\documentclass[%
reprint,
superscriptaddress,
nofootinbib,
 amsmath,amssymb,
 aps,
prc
]{revtex4-2}
\usepackage{graphicx}
\graphicspath{{./Images/}}
\usepackage{dcolumn}
\usepackage{bm}
\usepackage{slashed}
\usepackage{physics}
\usepackage{xfrac}
\usepackage{float}
\usepackage{color}
\usepackage[scr=boondox,  
            cal=esstix]   
           {mathalpha}
\usepackage[colorlinks = true,
            linkcolor = blue,
            urlcolor  = magenta,
            citecolor = blue,
            anchorcolor = blue]{hyperref}%

\newcommand{\f}[1]{/\!\!\!#1}

\newcommand{\p}{\overline{\mathcal{P}}}

\newcommand{\Sp}{\mathscr{p}}
\allowdisplaybreaks[1]

\begin{document}

\preprint{APS/123-QED}

\title{Modified Coherence and the Transverse Extent of Jets}

\author{Amit Kumar}
\affiliation{Department of Physics, University of Regina, Regina, SK  S4S 0A2, Canada}
\affiliation{Department of Physics and Astronomy, Wayne State University, Detroit, MI 48201.}

\author{Abhijit~Majumder} 
\affiliation{Department of Physics and Astronomy, Wayne State University, Detroit, MI 48201.}

\author{Chathuranga~Sirimanna}
\affiliation{Department of Physics, Duke University, Durham, NC 27708.}
\affiliation{Department of Physics and Astronomy, Wayne State University, Detroit, MI 48201.}

\author{Yasuki Tachibana}
\affiliation{Akita International University, Yuwa, Akita-city 010-1292, Japan.}

\date{\today}

\begin{abstract}
We present a study of the transverse size of parton showers and their diminishing interaction with the medium in the high virtuality stage of jet evolution. We consider the process of a hard quark produced in deep inelastic scattering off a large nucleus. Single gluon radiation from this quark, in the absence of scattering, is re-derived using wave-packets. This allows for a derivation of the quantum uncertainty size of the hard quark, at the point of splitting. This uncertainty size is then incorporated within a Monte-Carlo shower routine yielding transverse shower sizes noticeably larger than the classical antenna size of the shower. No clear relation is found between the full uncertainty size of the shower and the virtuality of the originating parton. The single gluon emission from the hard quark is then re-analysed for the case of single rescattering off the remainder of the nucleus. A relation is derived between the jet transport coefficient $\hat{q}$ and the gluon Transverse Momentum Dependent Parton Distribution Function (gTMDPDF). Solving this relation, for a simple case, clearly demonstrates the weakening of $\hat{q}$ with the virtuality of the hard splitting parton.
\end{abstract}


\maketitle



\section{Introduction}
\label{sec:Introduction}
The modification of high transverse momentum (high $p_T$) jets~\cite{ATLAS:2012tjt,CMS:2011iwn,ALICE:2013dpt} and hadrons~\cite{ATLAS:2015qmb,CMS:2012aa,ALICE:2010yje,STAR:2003fka,PHENIX:2001hpc} produced in a heavy-ion collision has become a central observable in the study of the quark-gluon-plasma (QGP) created in these collisions~\cite{Majumder:2010qh,Cao:2020wlm}. 
What started as the study of stimulated gluon radiation off hard partons~\cite{Wang:1991xy,Gyulassy:1993hr,Baier:1994bd} has now turned into a study of the modification of jets of arbitrary number of partons, using sophisticated event generators~\cite{Cao:2017qpx,Majumder:2013re, Zapp:2008gi,Zapp:2012ak,He:2015pra,Schenke:2009gb,Casalderrey-Solana:2014bpa}. The plurality of these generators has necessitated the use of event generator frameworks~\cite{Putschke:2019yrg}, aggregating multiple generators in modular fashion within an end-to-end simulation~\cite{Cao:2024pxc}.

Each of these event generators are based on different formalisms of energy loss~\cite{Guo:2000nz,Wang:2001ifa,Majumder:2009zu, Arnold:2002ja, Liu:2006nn, Majumder:2007iu}, each of which apply at a specific epoch of the evolution of the jet in the medium~\cite{Majumder:2010qh,Cao:2024pxc}. Jets start as single partons emanating from a hard collision. These partons are highly virtual, i.e., far off their mass shell. They undergo sequential decay into less virtual partons, each with opposing transverse momentum relative to the direction of the parent parton, producing a spray of partons very close to their mass shell. Throughout this process, the strong coupling grows with every split as the momentum scale depreciates. Portions of the jet, where the momentum scale begins to approach $\Lambda_{\mathrm{QCD}}$, will begin to interact strongly with the medium, if within it, or hadronize, eventually turning the entire partonic system into a jet of hadrons. 

Even within the purely perturbative sector of the jet, where the momentum scale is still considerably larger than $\Lambda_{\mathrm{QCD}}$, there are at least two different stages of jet modification~\cite{Cao:2017zih}, one where the effect of the medium is a perturbative correction to the vacuum shower~\cite{Guo:2000nz,Wang:2001ifa,Sirimanna:2021sqx}, and one where the medium induced radiation rate far exceeds the rate for a vacuum like split with the same kinematics~\cite{Majumder:2009ge,Arnold:2002ja,Baier:1996kr,Zakharov_1996}. 
In the first stage, the virtuality of the parent parton $\mu^2$, or the transverse momentum of the radiated parton $\ell_\perp^2 = \mu^2 y(1-y) \sim \mu^2 $ (where $y$ is the forward momentum fraction of one of the offspring), is much larger than the transverse momentum acquired by multiple scattering of the parent parton $\mu_{\mathrm{med}}^2 = \hat{q} \tau$, namely $\mu^2 \gg \mu_{\mathrm{med}}^2$~\cite{Majumder:2014gda}. 
The transport coefficient $\hat{q}$~\cite{Baier:2002tc,Majumder:2012sh,Kumar:2020wvb} is the mean squared transverse momentum acquired by the parton per unit length and $\tau$ is the formation time (equal to the formation length for light-like particles) given by
\begin{align}
    \tau = \frac{2E y(1-y)}{\ell_\perp^2} = \frac{2E}{\mu^2}.
\end{align}
In the second stage, the multiple scattering and rare emissions on average maintain the virtuality at the medium scale, namely the acquired transverse momentum: $\mu^2 \approx \mu_{\mathrm{med}}^2 = \hat{q} \tau$. (In this paper, we will consider $y$ values to be in the range from $0.1\lesssim y  \lesssim 0.9$, and thus transverse momenta are of the order of virtuality $\mu \sim \ell_\perp$.)

Using the equation above, and setting $\mu^2 = \mu_{\mathrm{med}}^2$ we obtain an expression for the medium generated scale, 
\begin{align}
    \mu_{\mathrm{med}}^2 = \sqrt{2 \hat{q} E}.  \label{eq:mu_med2}
\end{align}
This is often referred to as the saturation scale of the medium. Recent simulations using such a two-stage framework have demonstrated the need for separate formalism for parton virtualities above and at this scale~\cite{JETSCAPE:2022jer,JETSCAPE:2022hcb,JETSCAPE:2023hqn,JETSCAPE:2024nkj} (or the need for new formalism that can appropriately address both regimes), with a gradual \emph{weakening} of the effective transport coefficient $\hat{q}$ above this saturation scale. In Refs.~\cite{JETSCAPE:2022jer,JETSCAPE:2022hcb,JETSCAPE:2023hqn,JETSCAPE:2024nkj} the \emph{weakening} factor was  virtuality dependent, i.e.,
\begin{align}
    \hat{q} (T,E,\mu^2) = f(\mu^2) \hat{q} (T,E).
\end{align}
where $\hat{q}(T,E)$ is the Hard-Thermal-Loop (HTL) effective theory expression from Ref.~\cite{Arnold:2008vd}. The factor $f(\mu^2)$  was approximated from an effective description of the medium using a QGP parton distribution function (QGP-PDF) in Ref.~\cite{Kumar:2019uvu}. In a wide ranging Bayesian analysis carried out in Ref.~\cite{JETSCAPE:2024cqe}, which compared a multistage simulation with all leading hadron and jet data, the $f(\mu^2)$ was parametrized. The Bayesian analysis determined $f(\mu^2)$ to be a factor that drops swifty with $\mu^2$.

Complementary to these phenomenological observations of a gradual weakening of the transport coefficient, with the virtuality of a given parton, are the ``coherence" based arguments of Refs.~\cite{Casalderrey-Solana:2012evi, Mehtar-Tani:2011hma,Armesto:2011ir, Casalderrey-Solana:2011ule}. These argue that splits, before or even after formation, cannot be resolved until the distance between the daughters becomes larger than the ``typical" resolution scale of medium gluons $1/\mu_{\mathrm{med}}\simeq(2\hat{q}E)^{-1/4}$. Thus, the effective transport coefficient should be reduced, as the gluons from the medium can only perceive the overall color charge of the produced antenna. In other words, parton showers with transverse size below the inverse saturation scale are almost unaffected by the medium, and are, thus, vacuum-like.

The description above, while simple to state, is difficult to visualize or understand. Let us assume that the medium is large and static~\cite{JETSCAPE:2017eso} and has a resolution scale of $1/\mu_{\mathrm{med}}=(2\hat{q}E)^{-1/4}$. Let us further imagine that a jet with a virtuality $\mu \gg \mu_{\rm med}$ starts to propagate and shower in this medium. By the time that two partons (or parton clusters) are separated to the scale $1/\mu_{\rm med}$, there may have been many emissions, and the color charges of the two partons may not be easily related. Even if the two partons (or parton clusters) can be resolved by the medium at a later time, the medium can no longer affect the original process of emission as that has happened in the past. Thus, in the coherence picture, the only emissions that can be affected by the medium are those where the partons are separated by $1/\mu_{\mathrm{med}}$ within their formation time.

The calculations of coherence typically involve, a quark-anti-quark antenna formed from a virtual photon. The calculation of stimulated emission is carried out using the BDMPS-Z formalism, which is typically only applicable for virtualities at or below $\mu_{\mathrm{med}}$. Also, the formation of the quark-anti-quark antenna, and subsequent gluon radiation, at scales above $\mu_{\mathrm{med}}$,  is considered to be unaffected by scattering in the medium. The entire antenna structure is formed by successive angle ordered vacuum like emission, where successive spilts happen at monotonically smaller angles. 
Based on these antenna arguments, it has been stated that jets with larger virtuality, which lead to antenna with larger transverse momentum, will tend to be resolved more (or earlier) by the medium, because of the larger size that these will grow to; and as a result, will experience greater quenching~\cite{Mehtar-Tani:2011hma,Casalderrey-Solana:2012evi}.

In this paper, we will demonstrate that the picture of jet modification as outlined by the antenna approach may not be entirely accurate. Partons are quantum mechanical objects and thus possess an uncertainty in their momentum, and a corresponding uncertainty in their location. In the two dimensions transverse to the direction of propagation of a parton, one may na\"{i}vely estimate the uncertainty of the transverse location of the off-spring as (in natural units with $\hbar=1$), 
\begin{equation}
    \delta z_\perp \sim \frac{1}{ \ell_\perp },
\end{equation}
where, $\ell_\perp$ is the transverse momentum of either off-spring, with respect to the parent. 
Treating their location uncertainty as a size, we can ascribe a transverse size to a developing shower beyond the classical antenna size of the shower. Quantum uncertainty will make the overall transverse size of the shower larger than the case of the classical antenna.

Partons with lower virtualities may, due to their quantum uncertainty size, compensate for the smaller antenna size of the ensuing shower. We demonstrate how one can extract the quantum uncertainty size in the case of a final state shower in deep-inelastic scattering (DIS). Due to these compensatory effects, the transverse size of the shower is not directly related to the virtuality of the originating parton. This will be demonstrated using a vacuum simulation, which, for the first time, will include the quantum uncertainty size. 

To study the interaction of these jets with an extended medium, we will consider the process of deep-inelastic scattering on a large nucleus (A-DIS), where the outgoing radiating quark will interact with the glue field of the remaining nucleons along its path. 
While the transverse size of jets will be found to have a weak dependence on the off-shellness of the originating partons, we will demonstrate that the coupling between the medium and a radiating parton does indeed weaken at larger parton virtualities, leading to a reduction in the rate of stimulated emission. The cause for this will be found to depend on several factors, one of which is the transverse size of the radiating parton. Other factors include the scale and momentum fraction dependence of the Transverse Momentum Dependent Parton distribution Function (TMDPDF) which controls the momentum of the gluon exchanged between the jet and the medium. Both the scale and momentum fraction of the exchanged gluon will be found to depend on the scale of the hard parton. 

Our calculations will be carried out for a hard quark radiating a hard gluon in the higher twist formalism~\cite{Guo:2000nz,Wang:2001ifa,Majumder:2009ge,Sirimanna:2021sqx}, which is applicable to partons with a virtuality much larger than the medium scale $\mu^2_{\rm med} = \hat{q} \tau$.
To be clear, the current paper is \emph{not} a re-derivation of coherence effects in the higher-twist formalism. It is a revaluation of the gradient expansion carried out in all higher-twist calculations. Unlike prior efforts, the momentum fraction and scale dependence of the soft matrix elements will be retained and evaluated in the final expressions, where the differing dependencies of the hard part and soft matrix elements on the separation scale $\mu^2 \sim \ell_\perp^2$, will lead to the weakening of the interaction of the jet with the medium.

The remainder of the paper is organized as follows: 
In Sec.~\ref{sec:size_of_vacuum_emissions}, we re-introduce the reader to the formalism of DIS on a large nucleus using wave-packets, and consider the case of a single gluon emission from the struck quark, in vacuum. 
Here we re-organize the derivation to extract the transverse size from quantum uncertainty arguments. In Sec.~\ref{sec:MC}, we modify the MATTER generator in vacuum to include transverse uncertainty size of partons, and use this to estimate the transverse size of entire jets. This section will demonstrate that the transverse size of jets is wider than the antenna size, and does not have a strong dependence on the originating parton's virtuality. In Sec.~\ref{sec:Kernel}, we re-analyze the single gluon emission kernel at next-to-leading twist and derive a new relation between $\hat{q}$ and the gluon TMDPDF within a nucleon beyond the struck quark. In Sec.~\ref{sec:Results}, we use this new relation to derive the scale dependence of $\hat{q}$, and its weakening with the off-shellness of the radiating quark. In Sec.~\ref{sec:Summary}, we summarize and provide an outlook on extensions and some consequences of this effort.

\section{The size of Vacuum like emissions}
\label{sec:size_of_vacuum_emissions}

Hard jets (with energies $ \gtrsim 100$~GeV) start with an off-shellness that is orders of magnitude higher than typical scales in the medium and thus produce multiple vacuum like emissions prior to interacting strongly with the medium. To study this vacuum stage and the start of interactions with the medium, we consider the higher-twist approach which focuses on emissions that are vacuum like,  and, in addition, includes the interference terms between vacuum and medium induced emissions~\cite{Guo:2000nz,Wang:2001ifa,Majumder:2009ge,Sirimanna:2021sqx}. 

\begin{figure}
    \centering
    \includegraphics[width=0.4\textwidth]{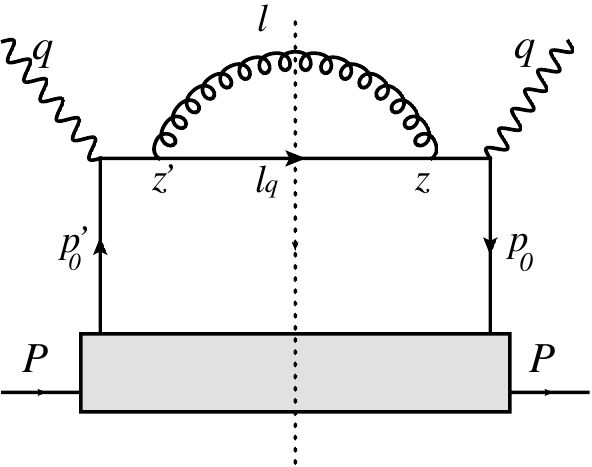}
    \caption{Single gluon emission from the outgoing struck quark produced in the deep-inelastic scattering off a large nucleus. The diagram above ignores final state interaction of the developing radiation. }
    \label{fig:basic-diagram}
\end{figure}

As is typically the case in higher twist calculations, we consider the process of a hard quark produced in the deep-inelastic scattering (DIS) off a large nucleus (A-DIS). For cases which ignore the correlation between initial state and final state interaction, as we are about to do in this paper, A-DIS provides a well controlled environment to study the produced jet. For the case of the first vacuum like emissions, we will not consider interactions in this section and will focus on hard gluon emissions from the produced quark (see Fig.~\ref{fig:basic-diagram}). 

Following Ref.~\cite{Majumder:2013re}, the differential cross section to produce a quark and gluon in the final state from the hard scattering of a lepton on a proton (or a nucleus), via one photon exchange, is written as, 

\begin{eqnarray}
\frac{ d \sigma  }{ d^{3} L d^{3} l  d^{3} l_q   } = \frac{\alpha_{\rm EM}^{2} \mathcal{L}^{ \mu \nu } }{2 \pi s E_{L} Q^{4}} 
\frac{ d \mathcal{W}_{ \mu \nu } }{ d^{3}  l  d^{3} l_q } ,
\label{eqn:diffcross}
\end{eqnarray}
where, $L$, $l$ and $l_q$ are the momenta of the recoiling lepton (with energy $E_L$), radiated gluon and out-going quark. The incoming proton (or nucleon in a nucleus) has a momentum of $P$, while the momentum of the photon is 
\begin{eqnarray}
    q = \left[  \frac{-Q^2}{2q^-} , q^- , 0, 0       \right].
\end{eqnarray} 
In Eq.~\eqref{eqn:diffcross}, the leptonic tensor $\mathcal{L}^{ \mu \nu } $ contains all portions of the Dirac trace of leptonic variables. This is connected by a single exchanged photon with the hadronic tensor $\mathcal{W}_{\mu \nu}$, which contains all hadronic fields:
\begin{eqnarray}
   \mathcal{W}^{ \mu \nu } \!=\! \frac{Disc.}{2\pi i \Omega} \left[ \int\!\! d^4 y d^4 y' e^{i q \cdot (y - y') } \langle P | J^{\mu} (y) J^\nu (y')  | P \rangle \right]\!.
\end{eqnarray}

In the equation above, $|P \rangle$ is the incoming hadronic state. Unlike typical expressions for $\mathcal{W}^{\mu \nu}$, we include space-time integrations over the locations of both currents $y$, $y'$. The extra space-time integration necessitates division by the four volume $\Omega$.
Using the collinear approximation $q^-,q^+ \gg k_\perp$, where $k_\perp$ represents a typical transverse momentum in the problem, we can factorize the hard part from the initial and final state of the process.

To understand the transverse size of the radiating antenna, we consider the process of a single emitted gluon in the final state. For this case, we obtain the hadronic tensor as, 
\begin{eqnarray}
    \mathcal{W}^{\mu \nu} \!\!=\! \frac{1}{\Omega} \!\!\int\! d^4 y d^4 y' {\rm Tr} \left[ \frac{\gamma^-}{2} \gamma^\mu \frac{\gamma^+}{2} \gamma^\nu \right] F(y,y') \mathcal{O} (y,y'), \; \label{eq:Wmunu-general}
\end{eqnarray}
where,
\begin{eqnarray}
    F(y,y') = \langle P | \Bar{\psi} (y) \frac{\gamma^+}{2} \psi(y') | P \rangle\; , \label{F-general}
\end{eqnarray}
and,
\begin{eqnarray}
    \mathcal{O}(y,y') &=& \int \frac{d^4 l}{(2\pi)^4} d^4 z d^4 z' \frac{d^4 l_q}{(2\pi)^4} \frac{d^4 p}{(2\pi)^4} \frac{d^4 p'}{(2\pi)^4}  \label{eq:O-general} \\
    &\times& g^2 {\rm Tr} \left[ \frac{\gamma^-}{2} \frac{  ( \f{p} + \f{q} )}{(p+q)^2 - i \epsilon}  \gamma^\alpha \f{l}_q  2 \pi \delta (l_q^2) \gamma^\beta \right. \nonumber \\  
    &\times& \left. \frac{  ( \f{p'} + \f{q} )}{(p'+q)^2 + i \epsilon} \right] 
e^{i q \cdot ( y - y')} e^{-i(p+q)\cdot (y -z)  } \nonumber \\
&\times& e^{-i (l+l_q) \cdot (z - z')} e^{i(p'+q)\cdot(y' - z')} G_{\alpha \beta} (l) 2\pi \delta(l^2)
. \nonumber
\end{eqnarray}

\subsection{Incoming wave-packets}

We can now introduce uncertainty in the incoming momentum states, by replacing the proton states $|P\rangle$ with single quark states, with transverse momentum distributed according to a Gaussian distribution:
\begin{align}
    | P \rangle &\longrightarrow& \int \frac{d^3 \Sp'}{(2\pi)^3} \frac{ \exp[-\frac{{\Sp'}_\perp^2}{2\Delta_\perp^2}   -\frac{ \left( \Sp'^+ - \p\mbox{}^+ \right)^2}{2\Delta_{+}^2}  ]}{\sqrt{2{\Sp'}^+}} | \Sp' \rangle. \label{eq:input-wave-function}
\end{align}
In the above equation, $\p\mbox{}^+$ is the large mean forward momentum of the incoming state. We assume that $\p\mbox{}^+$ is larger than all other scales in the problem. As a result, 
${\Sp'}^- \simeq {\Sp'}_\perp^2/( 2 { \Bar{\Sp'} }^+) $, and $\Delta_\perp^2$, $\Delta_+^2$ are widths that, we assume,
can be controlled by the experiment. 

The incoming function $F(y,y')$ can be re-expressed in terms of mean and uncertainty in $\Sp,y$ as, 
\begin{eqnarray}
    F(y,y') &=& \int \frac{d^3 \Sp}{(2\pi)^3 \sqrt{2\Sp^+}} \frac{d^3 \Sp'}{(2\pi)^3 \sqrt{2\Sp'^+}}  \label{eq:Fyy-expanded}\\
    &\times& \Bar{u}(\Sp) \frac{\gamma^+}{2} u(\Sp') e^{i\Sp\cdot y} e^{-i\Sp' \cdot y'}  e^\frac{-\Sp_\perp^2 - {\Sp'}_\perp^2}{2 \Delta_\perp^2} \nonumber \\
    &\times& e^{ - \frac{(\Sp^+ - \p\mbox{}^+ )^2 }{2\Delta_+^2  } }  
    e^{ - \frac{({\Sp'}^+ - \p\mbox{}^+ )^2 }{2\Delta_+^2  } } \nonumber \\
    &=& \int \frac{d^3 \Bar{\Sp}  d^3 \delta \Sp  }{(2\pi)^6  }  
     \frac{ \Bar{u}( \Bar{\Sp} + \frac{\delta \Sp}{2} ) \frac{\gamma^+}{2} u( \Bar{\Sp} - \frac{\delta \Sp}{2} )  }{ 2\sqrt{ {(\bar{\Sp}^+)}^2  - {(\delta \Sp^+)}^2/4  }  }\nonumber \\
     &\times& \exp{-\frac{\Bar{\Sp}_\perp^2}{\Delta_\perp^2 } - \frac{\delta \Sp_\perp^2}{4 \Delta_\perp^2}}  
     \exp { -\frac{ (\bar{\Sp}^+ - \p\mbox{}^+ )^2}{\Delta_{+}^2 }  }  \nonumber \\ 
     &\times& \exp[\frac{- \delta {\Sp^+}^2 }{4 \Delta_+^2}] 
     e^{i\bar{\Sp}\cdot \delta y} e^{-i \delta \Sp \cdot \bar{y}} \nonumber.
\end{eqnarray}
We have decomposed $\Sp = \bar{\Sp} + \delta \Sp /2$, $\Sp' = \Bar{\Sp} - \delta \Sp/2 $, $y= \Bar{y} + \delta y /2  $ and 
$y' = \Bar{y} - \delta y/2$.  The Gaussian widths applied on the transverse momenta are assumed different from those applied on the longitudinal momenta. See Ref.~\cite{Majumder:2013re}, where widths on the transverse momenta were ignored.  
In the limit of large $\p\mbox{}^+ \gg \Delta_+, \Delta_\perp $,  we obtain,
\begin{equation}
    \Bar{u}\left(\bar{\Sp} + \frac{\delta \Sp}{2}\right) \frac{\gamma^+}{2} u\left(\bar{\Sp} - \frac{\delta \Sp}{2}\right) \simeq 2 \p\mbox{}^+.
\end{equation}
Shifting $\bar{\Sp}^+ = \bar{\Sp}^+ + \p\mbox{}^+$, and approximating that $\p\mbox{}^+$ is larger than all other momenta in the problem, we obtain, 
\begin{eqnarray}
    \Sp^- \!\!\!\!&=&\!\! \frac{ ( \bar{\Sp}_\perp + \delta \Sp_\perp/2 )^2 }{2 ( \bar{\Sp}^+ + \delta \Sp^+/2 ) } \simeq \frac{\bar{\Sp}_\perp^2 + \frac{\delta \Sp_\perp^2}{4}}{2 \p\mbox{}^+} + \frac{ \bar{\Sp}_\perp \cdot \delta \Sp_\perp }{2 \p\mbox{}^+} , \nonumber \\
    {\Sp'}^- \!\!\!\!&=&\!\! \frac{ ( \bar{\Sp}_\perp - \delta \Sp_\perp/2 )^2 }{2 ( \bar{\Sp}^+ - \delta \Sp^+/2 ) } \simeq \frac{\bar{\Sp}_\perp^2 + \frac{\delta \Sp_\perp^2}{4}}{2 \p\mbox{}^+} - \frac{ \bar{\Sp}_\perp \cdot \delta \Sp_\perp }{2 \p\mbox{}^+}.
\end{eqnarray}
As a result, we obtain the mean and difference in the small $(-)$-component as: 
\begin{equation}
    \bar{\Sp}^- = \frac{(\Bar{\Sp}_\bot^2 + \sfrac{\delta \Sp_\bot^2}{4})}{2\p\mbox{}^+} \: {\rm and }\: \delta \Sp^- = \frac{\bar{\Sp}_\bot \cdot \delta \Sp_\bot}
    {  \p\mbox{}^+} .
\end{equation}  

Both of these terms are suppressed by the hard scale $\p\mbox{}^+$. In the high energy limit, we would obtain exponentials of the form $\exp[ i \bar{\Sp}^- \delta y^+ ]$ and $\exp[ i \delta \Sp^- \Bar{y}^+]$. Thus, when completely included in the $\Bar{\Sp}_\bot$ and $\delta \Sp_\perp$ integrations, these terms will restrict $\Bar{y}^+$ and $\delta y^+$ to values smaller than $\sim \sfrac{2\p\mbox{}^+}{\bar{\Sp}^2_\perp}$. We will ignore these terms, in this first attempt to determine the uncertainty size of the incoming hard parton. Thus, we will only have distributions in the $-$ and $\bot$ components.

If we were to integrate out $\Bar{\Sp},\delta \Sp$, the incoming state factor [$F(\bar{y},\delta y)$] reduces to, 
\begin{align}
    F(\Bar{y},\delta y ) &= e^{i \p\mbox{}^+ \delta y^-} 
    N e^{- [ 2(\bar{y}^{-}) \Delta_+]^{2} } 
    e^{-[ (\delta y^-) \Delta_+]^2 } \nonumber \\ 
    & \times e^{-\delta y_\perp^2 \Delta_\bot^2} e^{-\Bar{y}_\perp^2 4\Delta_\bot^2}.
\end{align}
The factor of $N$ includes all normalizations. 
Thus, the incoming state is centered at $\Bar{y} = 0$ with an uncertainty of $1/[4\Delta_p^2]$ (where $p \in + , \perp$). By setting $\Delta_p \simeq \lambda Q$ (where $\lambda \ll 1$, however $\lambda Q \gg \Lambda_{\mathrm{QCD}}$), we localize the ($y^-,y_\perp$) extent of the incoming and outgoing hard quark to a size of $(y^-)^2, y_\perp^2 \simeq 1/(\lambda Q)^2$.

\subsection{Wave-packets at split}

In this paper, we are not interested in the distribution at $y$ but rather the distribution at $z$, the location of the split of the hard quark into a quark and a gluon. This part can be calculated under the constraint that $y'=0$, i.e., the calculation of the transverse size of the radiating parton, at the location of the split, does not require the mean and uncertainty of the initial hard scattering location to be specified. However, in the interest of continuity with the preceding sub-section, we will retain these locations in what follows. Substituting Eqs.~(\ref{eq:O-general},\ref{eq:Fyy-expanded}) in Eq.~\eqref{eq:Wmunu-general}, and carrying out the $y,y'$ integration sets $p= \Sp$ and $p' = \Sp'$. The remaining expression for the hadronic tensor reads, 
\begin{eqnarray}
  \mathcal{W}^{\mu \nu} &=& \frac{ {\rm Tr} \left[ \frac{\gamma^-}{2} 
  \gamma^\mu \frac{\gamma^+}{2} \gamma^\nu \right] }{ \Omega} \\
  &\times& \int \frac{d^4 l}{(2\pi)^4} d^4 z d^4 z' \frac{d^4 l_q}{(2\pi)^4} 
  \frac{d^3 \Bar{\Sp}  d^3 \delta \Sp  }{(2\pi)^6  } \nonumber \\
  &\times& g^2 {\rm Tr} \left[ \frac{\gamma^-}{2} 
  \frac{  ( \f{\Sp} + \f{q} )}{(\Sp+q)^2 - i \epsilon}  
  \gamma^\alpha \f{l}_q  2 \pi \delta (l_q^2) \gamma^\beta \right. \nonumber \\  
  &\times& \left. \frac{  ( \f{\Sp'} + \f{q} )}{(\Sp'+q)^2 + i \epsilon} \right] 
 e^{i(\Sp+q)\cdot z  } 
 \exp{-\frac{\Bar{\Sp}_\perp^2}{\Delta_\perp^2 } - \frac{\delta \Sp_\perp^2}{4\Delta_\perp^2}}   \nonumber \\
 &\times& \exp { -\frac{ (\bar{\Sp}^+ - \p\mbox{}^+ )^2}{\Delta_{+}^2 }  }  
\exp[\frac{- \delta {\Sp^+}^2 }{4 \Delta_+^2}] \nonumber \\
&\times& e^{-i (l+l_q) \cdot (z - z')} e^{-i(\Sp'+q)\cdot z'} G_{\alpha \beta} (l) 2\pi \delta(l^2). \nonumber
\end{eqnarray}
Invoking the two on-shell delta functions, we get,
\begin{align}
    l^+ &= \frac{l_\perp^2}{2 l^-}, \,\,\, {\rm and} \,\,\, l_q^+ = \frac{{l_q}_\perp^2}{2 l_q^-}.
\end{align}
Once again, we invoke the shift to remove the large component of $\bar{\Sp}^+$ as $\bar{\Sp}^+ \rightarrow \bar{\Sp}^+ + \p\mbox{}^+$, and separate out each of the phase factors for sequential integration, 
\begin{eqnarray}
  \mathcal{W}^{\mu \nu} &=& \!\frac{ {\rm Tr} \left[ \frac{\gamma^-}{2} 
  \gamma^\mu \frac{\gamma^+}{2} \gamma^\nu \right] }{ \Omega} \label{eq:Wmunu-with-y-y'-integrated}\\
  &\times& \!\int \frac{d l^- d^2 l_\perp}{(2\pi)^3 2 l^-} d^4 \bar{z} d^4 \delta z 
  \frac{d l_q^- d^2 {l_q}_\perp}{(2\pi)^3 2 l_q^-} 
  \frac{d^3 \Bar{\Sp}  d^3 \delta \Sp  }{(2\pi)^6  } g^2\nonumber \\
  &\times&  {\rm Tr} \left[ \frac{\gamma^-}{2} 
  \frac{   \gamma^+ }{ 2 ( \bar{\Sp}^+ + \frac{\delta \Sp^+}{2} + \p\mbox{}^+ + q^+ - i \epsilon) }  
  \gamma^\alpha \f{l}_q  \gamma^\beta \right. \nonumber \\  
  &\times& \!\left. \frac{  \gamma^+ }{2 (\bar{\Sp}^+ - \frac{\delta \Sp^+}{2} 
  +  \p\mbox{}^+ + q^+ + i \epsilon)} \right]  G_{\alpha \beta} (l) \nonumber \\
 &\times& \exp{-\frac{\Bar{\Sp}_\perp^2}{\Delta_\perp^2 }  - \frac{\delta \Sp_\perp^2}{4\Delta_\perp^2}}   \nonumber \\
 &\times& \!\exp { -\frac{ (\bar{\Sp}^+)^2}{\Delta_{+}^2 }    
- \frac{ \delta {\Sp^+}^2 }{4 \Delta_+^2} }  \nonumber \\
&\times& \!e^{-i (l+l_q) \cdot \delta z }  
e^{i(\bar{\Sp}\cdot \delta z + \delta \Sp \cdot \bar{z} )  } 
e^{i(\p + q)\cdot \delta z} . \nonumber
\end{eqnarray}
In the equation above, $\p\mbox{}^- = \p_\perp = 0$.

We can carry out the integration of $\bar{z}$, as we are only interested  in the size given by $\delta z$.
One should note that there are 4 $\bar{z}$ integrations but only 3 independent components of $\delta \Sp$. As a result, the $\int d\bar{z}^+$ will produce a factor or length $L^+$. In Ref.~\cite{Majumder:2013re}, the focus was on the longitudinal location of the split. Hence, in that effort, all positions other than $\bar{z}^-$ were integrated over. In this effort, we are focused on the transverse size at the formation time. As such, we will integrate over all components of $\bar{z}$. The naive integrations over $\delta z$ would typically yield the energy momentum conservation conditions: 
\begin{equation}
    (2\pi)^4 \delta^4( \bar{\Sp} + \p + q - l - l_q).
\end{equation}
However, at this point, we will not integrate over the offset in transverse positions, as that will yield the transverse size. We can integrate over $\delta z^+$ and $\delta z^-$, yielding, 
\begin{eqnarray}
    (2\pi)^2 \delta (\bar{\Sp}^+ + \p\mbox{}^+ + q^+ - l^+ - l_q^+) 
    \delta (q^- - l^- - l_q^- ) .
\end{eqnarray}

Using the above equations, we can carry out the $l_q^-$ and $\bar{\Sp}^+$ integrations. All these simplifications yield, 
\begin{eqnarray}
  \mathcal{W}^{\mu \nu} &=& \frac{ {\rm Tr} \left[ \frac{\gamma^-}{2} 
  \gamma^\mu \frac{\gamma^+}{2} \gamma^\nu \right] }{ \Omega} L^+ \label{eq:Wmunu-without-delta-functions}\\
  &\times& \!\int \frac{d y d^2 l_\perp}{(2\pi)^3 2 y} d^2 \delta z_\perp
  \frac{ d^2 {l_q}_\perp}{(2\pi)^2 2 q^-(1-y) } 
  \frac{d^2 \Bar{\Sp}_\perp   }{(2\pi)^2  } \nonumber \\
  &\times& g^2 {\rm Tr} \left[ \frac{\gamma^-}{2} 
  \frac{   \gamma^+ q^- }{ \sfrac{l_\perp^2}{y} + \sfrac{ {l_q}_\perp^2 }{(1-y)} }  
  \gamma^\alpha \f{l}_q  \gamma^\beta \right. \nonumber \\  
  &\times& \!\left. \frac{  \gamma^+ q^-}{  \sfrac{l_\perp^2}{y} + \sfrac{ {l_q}_\perp^2 }{(1-y)} } \right] 
 \exp{-\frac{\Bar{\Sp}_\perp^2}{\Delta_\perp^2 } }    \nonumber \\
&\times& \exp { -\frac{ ( \sfrac{l_\perp^2}{2l^-} + \sfrac{{l_q}_\perp^2}{2 l_q^-} - \p\mbox{}^+ - q^+ )^2}{\Delta_{+}^2 } } \nonumber \\
 &\times&    
 G_{\alpha \beta} (l) 
e^{-i (\vec{l}_\perp + \vec{l_q}_\perp) \cdot \delta \vec{z}_\perp }.     \nonumber
\end{eqnarray}
In the equation above, the factor $\exp{ - i \vec{\bar{\Sp}}_\perp \cdot \delta \vec{z}_\perp }$ is ignored in comparison to $\exp{ i (\vec{l}_\perp + \vec{l_q}_\perp) \cdot \delta \vec{z}_\perp }$.  
This is based on a choice that $\bar{\Sp}_\perp$ has a narrower distribution compared to $l_\perp$ and 
${l_q}_\perp$ [controlled by our choice of $\Delta_\perp$ in the incoming wave-function in Eq.~\eqref{eq:input-wave-function}]. In the absence of this assumption, completing the $\delta \vec{z}_\perp$ integration will set $\vec{\mathscr{p}}_\perp = \vec{l}_\perp + \vec{l}_{q \perp}$. This will yield a Gaussian factor of $\exp { - \sfrac{ (\vec{l}_\perp +  \vec{l_q}_\perp)^2}{\Delta_\perp^2} }$. This implies that that in general $\vec{l}_\perp \neq -\vec{l_q}_\perp$.  Assuming that $l_\perp, {l_q}_\perp \gg \Delta_\perp$, we would obtain the simple relation $\vec{l}_\perp \simeq -\vec{l_q}_\perp = \vec{\ell}_\perp $.

While the exact distribution of $\delta z_\perp$ may be difficult to surmise from the above form, we can calculate the leading contributions to the width of this distribution, by calculating, 
\begin{eqnarray}
    \langle \delta z_\perp^2 \rangle = \frac{ \int d^2 \delta z_\perp  \delta z_\perp^2 \frac{\partial^2 \mathcal{W}\cdot g_\perp } {d^2 \delta z_\perp } } { \mathcal{W}\cdot g_\perp }. \label{eq:delta-z-perp-definition}
\end{eqnarray}
Where, $\mathcal{W}\cdot g_\perp = \mathcal{W}_{\mu \nu} g_\perp^{\mu \nu}$.
Carrying out the $\Sp_\perp$ integration, the above expression can be expressed as, 
\begin{align}
\langle \delta z_\perp^2 \rangle &=\!\! \int \!\!d^2 \delta z_\perp dy d^2 l_\perp d^2 {l_q}_\perp  
\frac{ g_\perp^{\mu \nu}{\rm Tr} \left[ \frac{\gamma^-}{2} 
  \gamma_\mu \frac{\gamma^+}{2} \gamma_\nu \right] L^+ \Delta_\perp^2 }{ 2 \Omega  (2\pi)^7} 
  \nonumber \\
  &\times \left[  (-1) \vec{\nabla}_{l_\perp} \!\! \cdot \vec{\nabla}_{{l_q}_\perp} \!\!
  \frac{g^2 G_{\alpha \beta} (l)}{2q^- y(1-y)}  
\!\frac{ {\rm Tr} [ \gamma^+  \gamma^\alpha \f l_q \gamma^\beta ] }{4 L_\perp^2} \right. \nonumber\\
  &\times \left.  
  e^{ - \frac{ ( L_\perp - \p\mbox{}^+ - q^+)^2 }{\Delta_+^2 } } \right] 
  e^{-i (\vec{l}_\perp + \vec{l_q}_\perp) \cdot \delta \vec{z}_\perp }\Bigg/ \mathcal{W}\cdot g_\perp. 
  \label{eq:delta-z-perp-general-expression}
\end{align}
In the equation above, $L_\perp = \sfrac{l_\perp^2}{[2q^- y]} + \sfrac{ {l_q}_\perp^2 }{[2q^- (1-y)]} $. Thus, for either gradient $\nabla_{l_\perp}$, or $\nabla_{{l_q}_\perp}$, we obtain, 
\begin{align}
    \vec{\nabla}_{l_\perp} \mathscr{f}( L_\perp ) = \frac{ \partial \mathscr{f} }{\partial L_\perp} \frac{ \partial L_\perp}{ \partial l_\perp^2} 2 \vec{l}_\perp,  \label{eq:partial-lperp}
\end{align}
where, $\mathscr{f} = \exp{ - \sfrac{ ( L_\perp - \p\mbox{}^+ - q^+)^2 }{\Delta_+^2 } } / L_\perp^2 $. 
The factor $\p\mbox{}^+ + q^+ $ is positive definite and is related to the mean virtuality of the hard quark after scattering as, 
\begin{equation}
  \langle\mu^2\rangle \simeq 2 (\p\mbox{}^+ + q^+) q^-.
\end{equation}
Using the above expression in combination with the mean formation time $\tau = \sfrac{2q^-}{\langle\mu^2\rangle} $, we can carry out the integrations over $\delta z_\perp$. The remaining term that depends on $l_\perp$ and ${l_q}_\perp$ is $\mathscr{g} = G_{\alpha \beta}(l) Tr [\gamma^+ \gamma^\alpha \f l_q \gamma^\beta ] $.

Carrying out the successive differentials, using Eq.~\eqref{eq:partial-lperp}, and setting $\vec{l}_\perp = -\vec{l_q}_\perp = \vec{\ell}_\perp $, will yield several terms. In this first attempt to derive the transverse width, we will only retain the most singular contribution in terms of $\ell_\perp$, i.e., terms which contain the quadratically divergent $\sfrac{d\ell_\perp^2}{\ell_\perp^4}$ terms.  Collecting these terms from 
$\vec{\nabla}_{l_\perp} \!\! \cdot \vec{\nabla}_{{l_q}_\perp} ( \mathscr{f\cdot g})$, we obtain, 
\begin{align}
\langle \delta z_\perp^2 \rangle &=  \int  dy d^2 \ell_\perp  \,
\frac{ 2 L^+ \Delta_\perp^2 8 g^2 }{  \Omega  (2\pi)^5} 
   \label{eq:delta-z-perp-simplified} \\
  &\times  \frac{ \bigg[ 3 (1-y)^3 + \frac{3}{2} (1-y) - 1 \bigg]}{ \ell_\perp^4} \nonumber \\
  & \times \exp{ -\frac{ \left( \sfrac{\ell_\perp^2}{[2q^- y(1-y)]} - 1/\tau \right)^2}
  {\Delta_+^2}  } \Bigg/ \mathcal{W}\cdot g_\perp. \nonumber
\end{align}

Substituting $\ell_\perp^2 = \mu^2 y(1-y)$, where $\mu^2$ is the virtuality of the parton before the split, we obtain, 
\begin{align}
\langle \delta z_\perp^2 \rangle &= \frac{ 2 L^+ \Delta_\perp^2 8 \alpha_S }{  \Omega  (2\pi)^3} 
 \int  dy d \mu^2 \exp{ -\frac{ \left( \mu^2 - \langle \mu^2 \rangle \right)^2 }{  ( 2 q^-\Delta_+) ^2}  }
  \nonumber \\
  &\times  \frac{ \bigg[ 3 (1-y)^3 + \frac{3}{2} (1-y) - 1 \bigg]}{ \mu^4 y (1-y)} 
  \Bigg/ \mathcal{W}\cdot g_\perp. \label{eq:delta-z-perp-y-dependent}
\end{align}
Carrying out the $y$ integration, we would obtain, 
\begin{align}
\langle \delta z_\perp^2 \rangle &=  \frac{ \frac{ 2 L^+ \Delta_\perp^2 \alpha_S }{  \Omega  (2\pi)^3} 
 \int \frac{d \mu^2}{\mu^2} \frac{20}{\mu^2} \log (\frac{1}{\epsilon}) 
 e^{ -\frac{ \left( \mu^2 - \langle \mu^2 \rangle \right)^2 }{  ( 2 q^-\Delta_+) ^2}  }
 }{ \mathcal{W}\cdot g_\perp }. \label{eq:delta-z-perp}
\end{align}
The integration over $y$ in Eq.~\eqref{eq:delta-z-perp-simplified} is carried out from $\epsilon$ to $1-\epsilon$, yielding $\int_\epsilon^{1-\epsilon}1/y \simeq \log(\sfrac{1}{\epsilon})$, where $\epsilon$ is an infrared regulator (the exact form of this regulator has no significance for the calculation of $\langle \delta z_\perp^2\rangle$). 
Using similar approximations and steps for the denominator, we obtain, 
\begin{align}
    \mathcal{W}\cdot g_\perp \!\!= \frac{ 2 L^+ \Delta_\perp^2 \alpha_S }{  \Omega  (2\pi)^3}  
\!\! \int \!\frac{d \mu^2}{\mu^2}  2\log (\frac{1}{\epsilon}) 
 e^{ -\frac{ \left( \mu^2 - \langle \mu^2 \rangle \right)^2 }{  ( 2 q^-\Delta_+) ^2}  }. \label{eq:Wmunu}
\end{align}
Inspection of Eqs.~(\ref{eq:delta-z-perp},\ref{eq:Wmunu}) yields the approximate relation between the size of a splitting parton and its virtuality: 
\begin{align}
    \delta z_\perp^2 \simeq \frac{10}{\mu^2} \simeq \frac{10 \langle y(1-y) \rangle }{\ell_\perp^2 }. \label{eq:z-with-mu}
\end{align}
In the equation above $y$ is the forward momentum fraction of the radiated gluon. The angle brackets $\langle y(1-y) \rangle$ indicate the mean or typical value of the $y$-dependent factor (since $y$ was integrated out to obtain the above equations). For a typical value of the momentum fraction of a gluon radiated from a quark ($y \sim \sfrac{1}{10}$), one obtains the expected relation: 
\begin{equation}
    \delta z_\perp \approx \frac{1}{\ell_\perp}, \label{eq:z-with-l_perp}
\end{equation}
in units where $\hbar = 1$ (in real units: $\delta z_\perp \approx \sfrac{\hbar}{\ell_\perp}$).

The equation above is what one would have expected from a \emph{na\"{i}ve} application of the uncertainty principle. 
It is reassuring to recover this relation using a fully quantal treatment of a hard quark radiating a gluon. 
We hasten to point out that, in the derivation above, the parent parton (quark) has a vanishing average transverse momentum. 
The hard quark, produced in the deep-inelastic scattering of the virtual photon with the incoming quark (Fig.~\ref{fig:basic-diagram}), proceeds along the outgoing ``jet axis". 
The radiating gluon and outgoing quark (which encounter the cut line in Fig.~\ref{fig:basic-diagram} and are thus considered on shell) possess opposing transverse momenta of $\vec{\ell}_\perp$. However, the uncertainty size of the \emph{parent quark} is $\sim \sfrac{1}{\ell_\perp}$. 
The parent quark reaches this size at the point where it splits into the outgoing quark and gluon [with momentum fractions $y$ and $(1-y)$], i.e., at the formation time $\tau = \sfrac{2 q^- y(1-y) }{\ell_\perp^2}$.

All of the above considerations apply to the single split from a hard quark. Realistic jets contain many partons which arise from multiple splits, each of which will be affected by its uncertainty size. 
In the subsequent section, we will estimate the combined effect of these multiple splits, each with a minimum uncertainty size, on the entire transverse size of jets.

\section{Calculating jet transverse size with a Monte-Carlo implementation}
\label{sec:MC}

In the preceding section, we derived the transverse uncertainty size of a radiating quark, at the point where the radiating gluon and outgoing quark have split off from the parent quark. 
This represents the size at one split. 
In a typical jet shower, there are several splits, with successive partons emitted at larger angles away from the jet direction. 
Thus, the outgoing shower spreads to a larger transverse size with increasing time.
The most straightforward means to calculate this increasing transverse size is to use a Monte-Carlo parton shower event generator.

All current Monte-Carlo simulators of parton showers in the 
vacuum~\cite{Sjostrand:2007gs, Marchesini:1991ch,Gleisberg:2008ta}, have no need to determine this size, and thus operate entirely in momentum space. All current Monte-Carlo simulators of partons showers in a medium~\cite{Majumder:2013re,Cao:2017qpx,Schenke:2009gb,He:2015pra,Cao:2016gvr,Zapp:2008gi,Casalderrey-Solana:2014bpa} tend to use either the classical (antenna) size~\cite{Mehtar-Tani:2011hma} of the shower, or do not ascribe any transverse size to the shower. In this section, we will outline the differences between the quantum transverse uncertainty size and the classical or antenna transverse size of the shower.

A related question is when does the radiating or splitting parent quark reach this transverse size. 
As stated above, this size is reached at the point of the split. 
The longitudinal location of the split depends on the formation time which describes the mean longitudinal location of the split. Returning to Eq.~\eqref{eq:Wmunu-with-y-y'-integrated}, we note that we had integrated over 
$\bar{z} \mbox{}^-$ yielding a $\delta(\delta p^+)$. 
This was done in the interest of keeping the expressions less cumbersome. 
Following Ref.~\cite{Majumder:2013re}, we can integrate over $\delta p^+$ instead to obtain a multiplicative Gaussian factor of $ \int d\bar{z}\mbox{}^- \exp{- \bar{z}\mbox{}^- \Delta_+^2 }$. This factor will remain in all following equations including Eqs.~(\ref{eq:delta-z-perp}) and (\ref{eq:Wmunu}). Incorporating this factor will modulate the entire form for $\mathcal{W}^{\mu \nu}$ in Eq.~\eqref{eq:Wmunu-without-delta-functions}, leading to a Gaussian dependence of the actual mean split time $\bar{z}^-$, as demonstrated in Ref.~\cite{Majumder:2013re}. As done in Ref.~\cite{Majumder:2013re}, one sets $\Delta_+$ so that the mean split time equals the formation time $\tau = 2E/\mu^2$, where $\mu$ is the off-shellness or virtuality of the splitting parton. 

\begin{figure}
    \centering
    \includegraphics[width=0.8\linewidth]{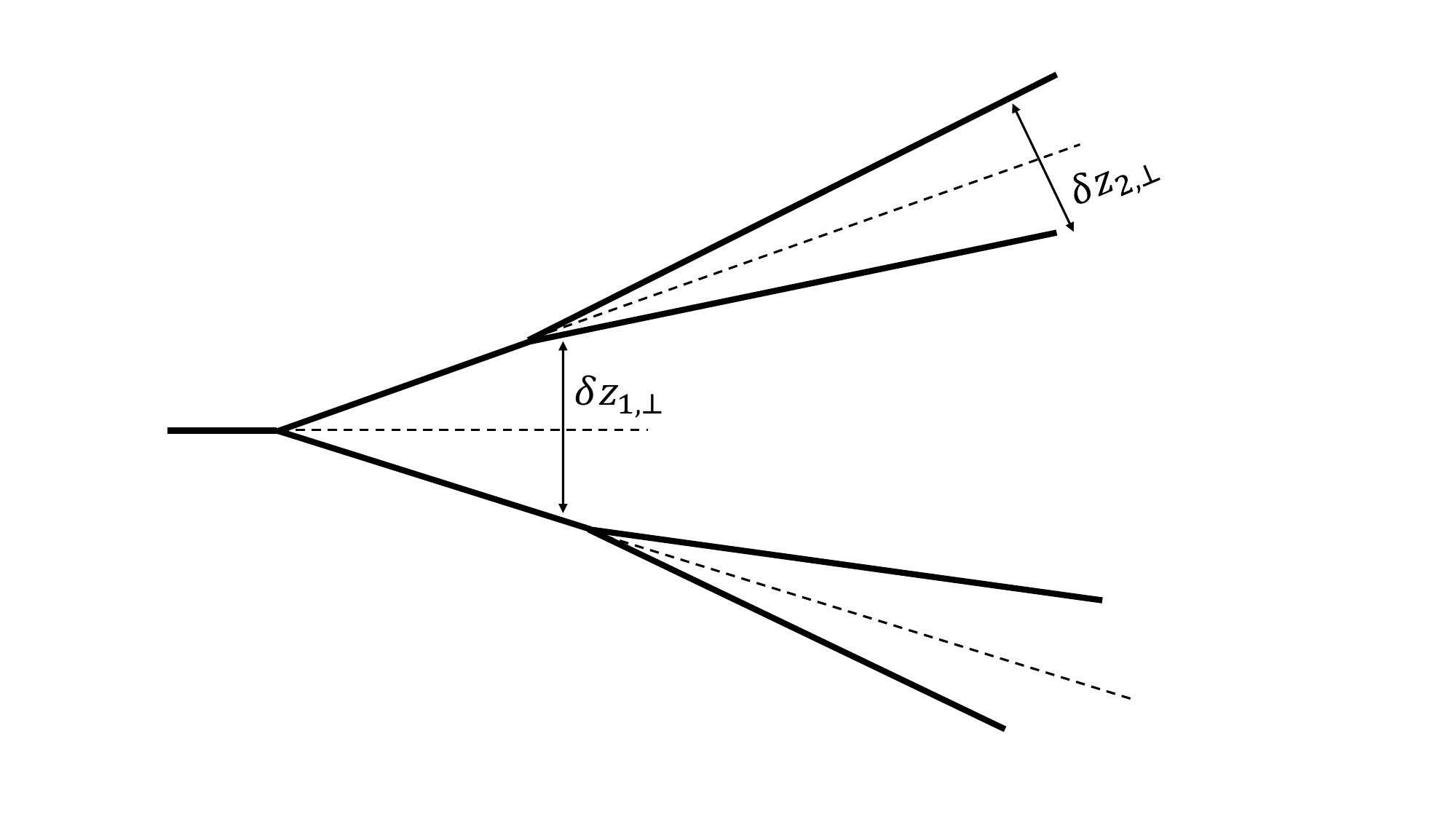}
    \caption{The antenna picture of a developing parton shower, in coordinate space. Most jet configurations are not planar (as seemingly indicated by the figure).}
    \label{fig:antenna-size}
\end{figure}

The uncertainty size $\delta z_\perp$ combined with the split location $\bar{z}\mbox{}^-$, allows for multiple interpretations of the transverse size of the radiating partons.  
The simplest would be the typical (or classical) antenna size: Assume partons are classical point like particles, that follow classical straight line trajectories. 
This case is illustrated in Fig.~\ref{fig:antenna-size}.
The initial partons have the largest virtuality, even compared to their energy, and thus have the shortest formation times, and the shortest lengths in the figure. The first split typically has the largest $l_\perp$ and thus is drawn with the largest angle. 
Subsequent partons tend to live longer (and thus have longer path-lengths), and smaller opening angles, with smaller values of transverse momentum $\ell_\perp$. 
It should be stressed that what is drawn in Fig.~\ref{fig:antenna-size} represents the typical event. 
Specific events may present with the later splits being wider than earlier spits, if angular ordering is not imposed. 

\begin{figure}
    \centering
    \includegraphics[width=1\linewidth]{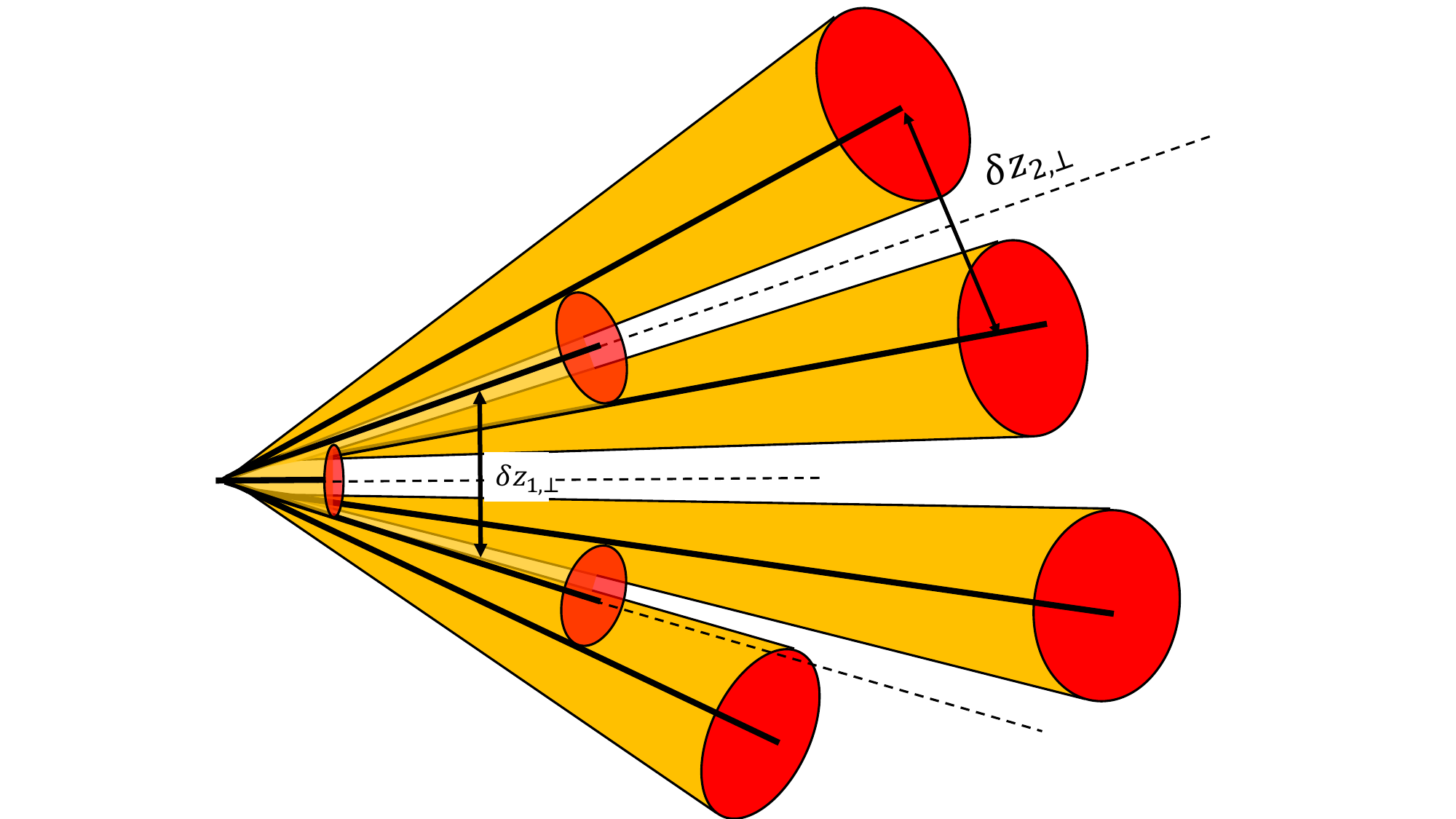}
    \caption{The picture of a developing shower, in coordinate space, using quantum uncertainty arguments. The partons within each wavepacket have the same momenta as in Fig.~\ref{fig:antenna-size}. }
    \label{fig:uncertainty-size}
\end{figure}

Consider the same splitting process, but now with the inclusion of quantum uncertainty.  Each splitting parton will have a transverse size given by Eq.~\eqref{eq:z-with-mu} or Eq.~\eqref{eq:z-with-l_perp}. 
This transverse size is indicated using the orange cones surrounding the partons in Fig.~\ref{fig:uncertainty-size}. The cones represent the mean uncertainty of the location of the split point. As a result, the outgoing partons are located on these cones. 
A ``semi-classical" means to imagine the developing wave-packets is to consider that each wave packet contains, within it, the two wave-packets of the partons that are about to split. 
These splitting wave-packets separate from each other at the formation length, at which point they are on the cone boundary, with their \emph{mean} location given by their respective momenta. 
The momenta of the partons in Fig.~\ref{fig:antenna-size} and Fig.~\ref{fig:uncertainty-size} are the same. However, the larger size of the shower due to quantum uncertainty, compared to the classical antenna picture, is evident. 

We point out here that the picture generated in Fig.~\ref{fig:uncertainty-size}, is one means to give a classical interpretation to a quantum phenomena. One could take a somewhat different approach, e.g., not limiting the cones to the mean width, but requiring both daughter partons to originate at the same point on the cones. While this may lead to a slightly different estimate of the transverse size of the shower, it \emph{will} lead to a shower that is wider than that given by a classical antenna simulation.

To obtain a quantitative estimate of this effect we calculate the average size of showers using the MATTER vacuum generator~\cite{Majumder:2013re,Cao:2017qpx} within the JETSCAPE framework~\cite{Putschke:2019yrg}. 
A single parton with fixed momentum along the $x$-axis is initiated and then allowed to shower for a fixed amount of time. 
To determine the size of the shower at a given time, we use the following measure:
\begin{align}
    r_{\perp} &= \frac{1}{N_{events}} \sum_{i=1}^{N_{events} } \frac{1}{N_i} \sum_{j=1}^{N_i} \sqrt{ y_j^2 + z_j^2 + \delta {r_j}_\perp^2 }. \label{eq:r_perp}
\end{align}
In the equation above, $N_i$ represents the number of partons in event $i$, and $y_j,z_j$ represent the $y,z$ components of parton $j$ at the time where the size is being determined. The $\delta {r_j}_\perp$ is the uncertainty size of the splitting parton at the time when the size is being determined. In the case of antenna simulations, i.e. Fig.~\ref{fig:antenna-size}, $\delta {r_j}_\perp = 0$. However, this is non-zero for the case of simulations where the quantum size is retained in the simulation (Fig.~\ref{fig:uncertainty-size}). 

Throughout this section, we use the term, ``transverse size of jet shower" (or ``transverse size of parton shower", or just simply ``transverse size of shower"), instead of ``transverse size of jet". This is because no jet clustering algorithm~\cite{Cacciari:2008gp} has been exercised on the final partons. We consider all the partons within some specified kinematic conditions, emanating from an initial hard quark.

The originating parton, a quark with an $E=100$~GeV starts at $x=y=z=0$, and moves in the $x$-direction. The virtuality is assigned by sampling the Sudakov form factor with a maximum allowed value of $Q\lesssim E/2$. This choice of maximum allowed virtuality has led to the agreement between JETSCAPE simulations and a wide range of experimental data~\cite{Kumar:2019bvr,JETSCAPE:2022jer,JETSCAPE:2022hcb,JETSCAPE:2023hqn,JETSCAPE:2024cqe}. 
As splits happen, the outgoing shower spreads in the $y$ and $z$ direction. With repeated splits, the virtuality of the partons continues to drop. A given parton stops splitting once its virtuality reaches $Q_0 \lesssim 1$~GeV. Partons that have a virtuality at or below $Q_0$,  or those that have a split time well in excess of the time of observation ($t=5$~fm/c), are free-streamed to this point in time. 

\begin{figure}
    \centering
    \includegraphics[width=\linewidth]{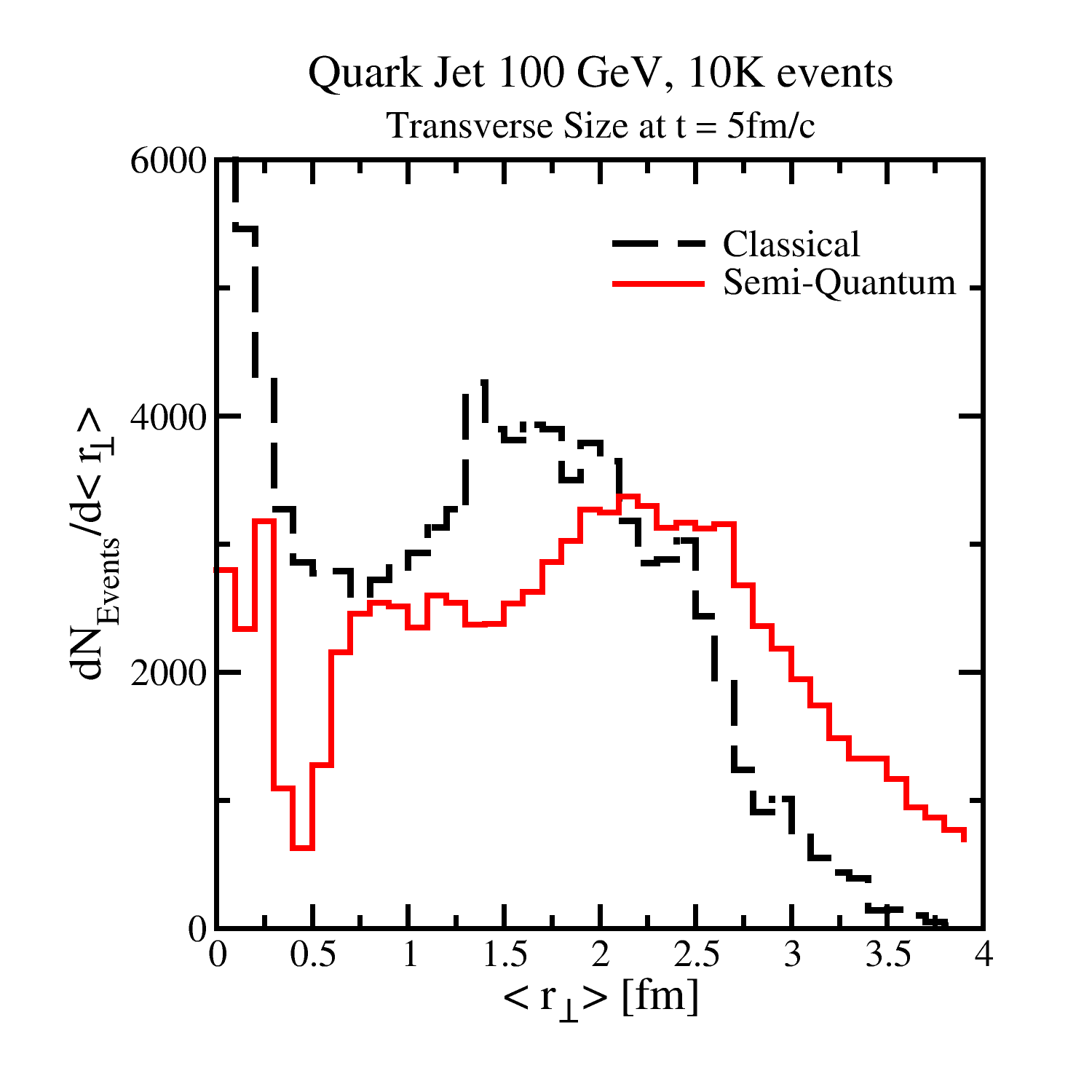}
    \caption{Comparison of the transverse size of a developing shower from a 100~GeV quark jet at 5fm/c. The black dashed line is the classical antenna size, while the red solid line in the quantum uncertainty size. The maximum allowed off-shellness of the initial hard parton is $Q \leq 50$~GeV. The plot clearly shows the larger quantum size compared to the classical antenna size. See text for further details. }
    \label{fig:classical_vs_quantum_size}
\end{figure}

\begin{figure}
     \centering
     \includegraphics[width=\linewidth]{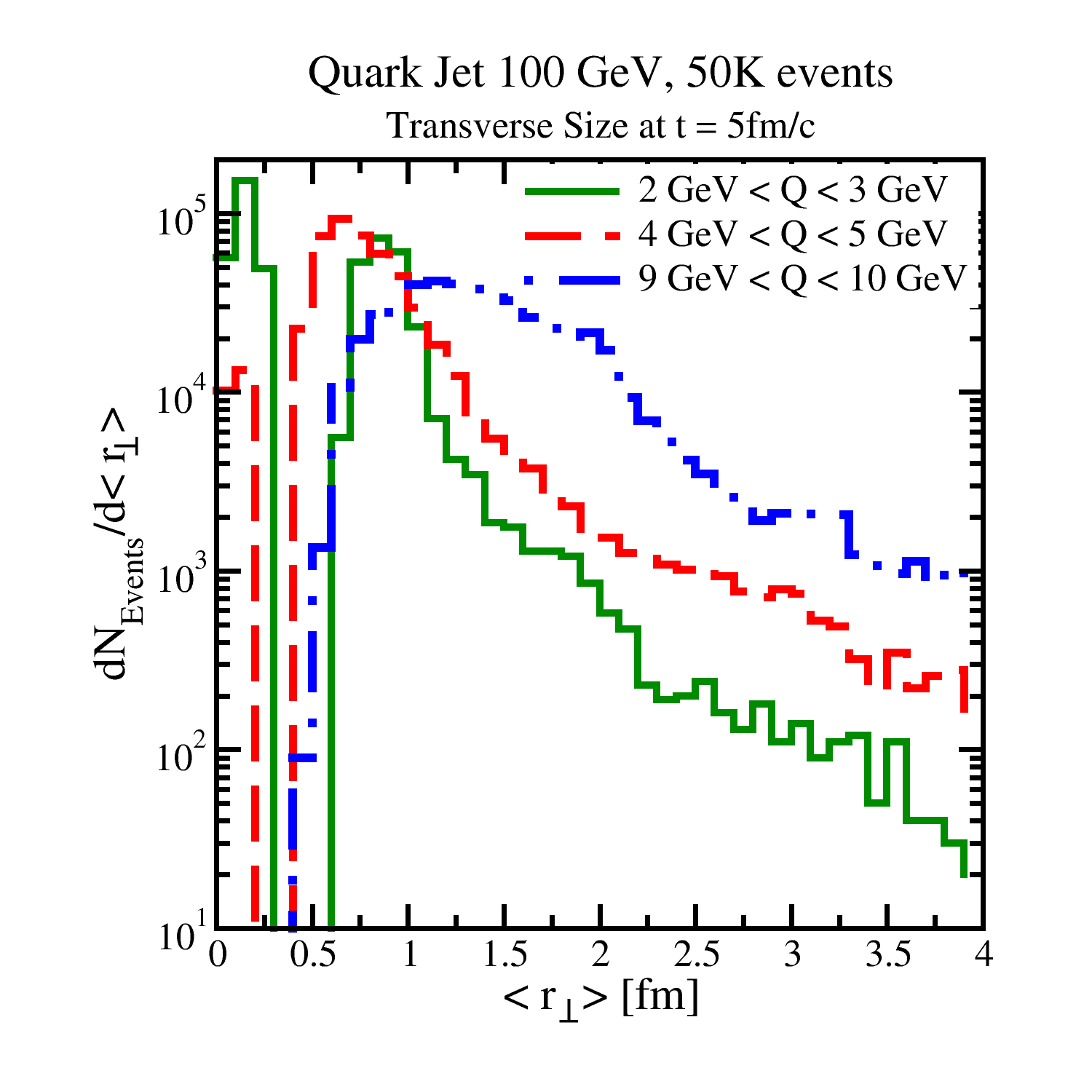}
     \caption{The quantum transverse size distributions of jets at $t=5$~fm/c, with initial off-shellness $(Q)$ limited to 3 narrow windows. The initial parton in each case is a quark with $E=100$~GeV. See text for description of plots.}
     \label{fig:quantum_size_vs_mass}
 \end{figure}

We first present the results of two separate simulations, each with 10K events: 
The first, where only the antenna size of the shower is 
retained in the simulation (this is what one would obtain by running MATTER~\emph{out-of-the-box}). The transverse size is calculated using partons in the jet's hemisphere: $p_x\geq 0$. 
This is indicated by the black dashed line in Fig.~\ref{fig:classical_vs_quantum_size}. 
The transverse size has a rising contribution at $r_\perp \rightarrow 0$. 
This emanates from contributions where the partons do not split (or have one soft split), and thus have an $r_\perp \simeq 0$. 
The entire distribution also has a wide and local maxima at $1.5 \lesssim r_\perp \lesssim 2$~fm. To our knowledge, this is the first paper that attempts to calculate the transverse size of a hard jet shower in coordinate space. The plot indicates that at 5~fm/c, where jets typically have their maximal interaction with the dense medium~\cite{JETSCAPE:2022jer} in a heavy-ion collision, the overall size of the jet shower is considerable, $\sim 2$~fm. This is just the effective radius of the jet shower as defined by Eq.~\eqref{eq:r_perp}. All simulations in this section are carried out in the vacuum, and no hadronization is applied. The sizes presented in Figs.~\ref{fig:classical_vs_quantum_size},~\ref{fig:quantum_size_vs_mass} are partonic sizes. 

In the second set of simulations, we initialize the hard parton in the same way as above; however, at each split, we shift the physical location of the daughter partons to the edge of the wave-packet, as indicated in Fig.~\ref{fig:uncertainty-size}. 
As would be expected, this leads to a larger transverse size of the jet shower at $t=5$~fm/c. This is indicated by the red solid line in Fig.~\ref{fig:classical_vs_quantum_size}. One notices the lack of the spike at $r_\perp \rightarrow 0$. This is because partons that do not split, now have an uncertainty size given by $\delta r_\perp$. 

The initial hard quark in either simulation, with or without uncertainty broadening, are restricted to a maximum virtuality of $Q\lesssim E/2$ ($E=100$~GeV, the energy of the parton).
In each event, the actual virtuality of the parton can vary, logarithmically, between $Q_0 \simeq 1$~GeV and $E/2$. 
One may infer that this wide range in virtuality yields the broad distribution of jet transverse sizes in Fig.~\ref{fig:classical_vs_quantum_size}.
This is not entirely true. 
While partons with very large initial virtuality do eventually lead to jets with a wide transverse size, at intermediate virtualities, the relation is not very clear cut. 
To illustrate this, we plot the transverse size of jets at $t=5$~fm/c, which start within narrow ranges of virtuality. The three cases studied in Fig.~\ref{fig:quantum_size_vs_mass}, are 2~GeV~$\leq Q \leq$~3~GeV (green solid), 4~GeV~$\leq Q \leq$~5~GeV (red dashed), and 9~GeV~$\leq Q \leq$~10~GeV (blue dot-dashed). 
The number of events is increased to 50K for better statistics.

While jet showers produced in the large virtuality bin (9~GeV$\leq Q \leq$10~GeV) do indeed produce larger transverse sizes, 
at lower virtualities, there seems to be no clear relation between virtuality and transverse size. 
This is due to the compensatory effect between the intrinsic uncertainty size of individual partons [$\delta {r_j}_\perp$ in Eq.~\eqref{eq:r_perp}], and the antenna size of the shower [$y_j,z_j$ in Eq.~\eqref{eq:r_perp}]. Lower virtualities lead to smaller antenna sizes, but larger intrinsic uncertainty sizes (and \emph{vice-versa}). 

Two observations stand out in the current section: 
As stated above, the relation between virtuality and the transverse size of jet showers is not straightforward. 
Second, jet showers have a much larger transverse size than what one would na\"{i}vely expect. 
They are not narrow phenomena localized along a line, but may interact with a large region of the medium (if they were produced in a heavy-ion collision, or other media).

\begin{figure}
    \centering
    \includegraphics[width=\linewidth]{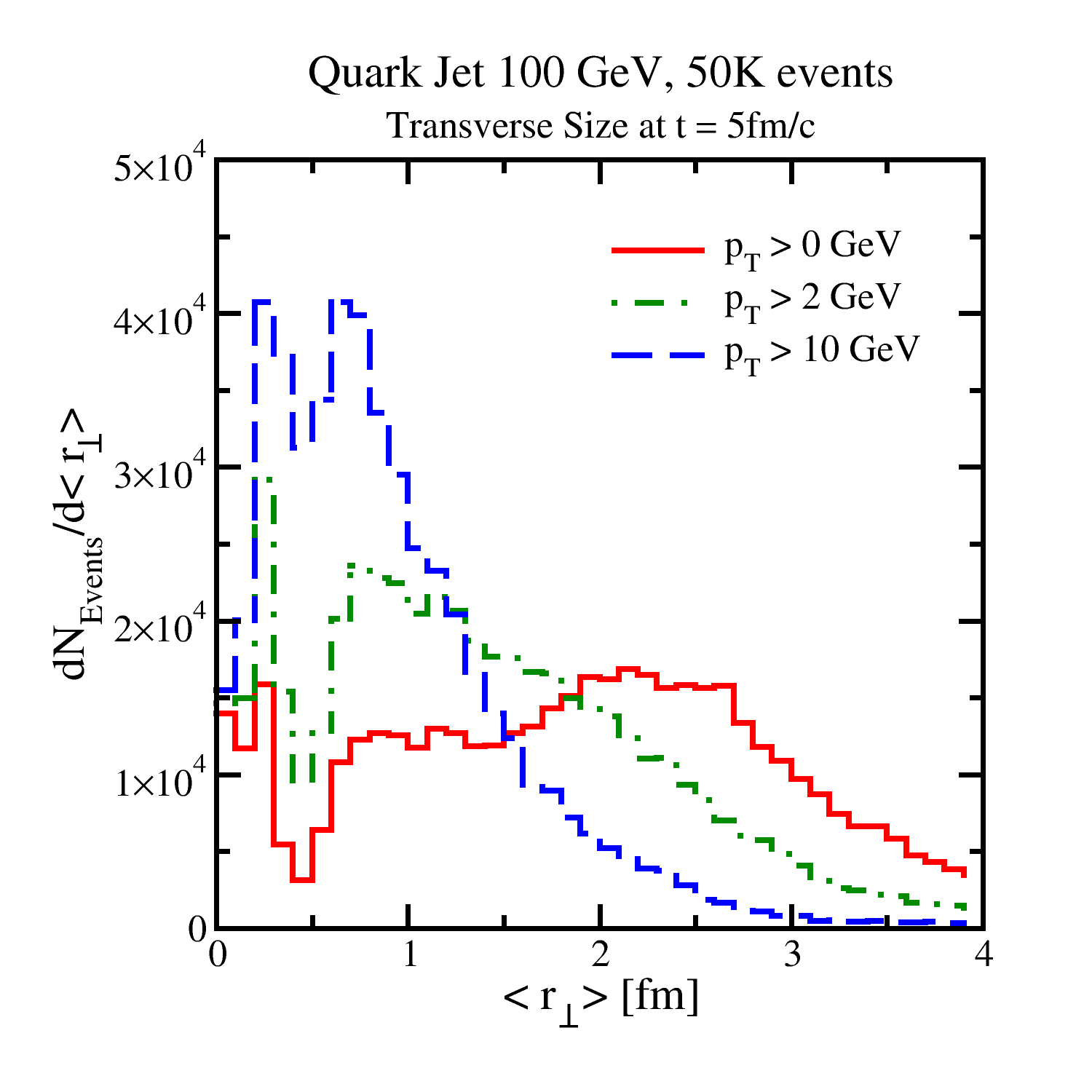}
    \caption{The transverse size of the jet shower with a minimum $p_T$ cutoff on the included partons. Results for $p_T > 0$~GeV, 2~GeV, or 10~GeV are plotted and show a narrowing of the jet shower with increasing $p_T$ of the constituents. }
    \label{fig:size_vs_pT}
\end{figure}

Also, the widest parts of the jet are composed of the softest partons, with the hard core propagating along a narrower cone. This is illustrated in Fig.~\ref{fig:size_vs_pT}. We again start with a quark, possessing an energy of $E=100$~GeV, with its virtuality distributed logarithmically from $Q_0 = 1$ to $Q_{\rm Max} = 50$~GeV. The solid red line illustrates the same transverse size distribution as the \emph{semi-quantum} line in Fig.~\ref{fig:classical_vs_quantum_size}. The green dot-dashed line results when we only retain partons with $p_x > 2$~GeV, and the blue dashed line only contains partons with $p_x > 10$~GeV.

\begin{figure}
    \centering
    \includegraphics[width=\linewidth]{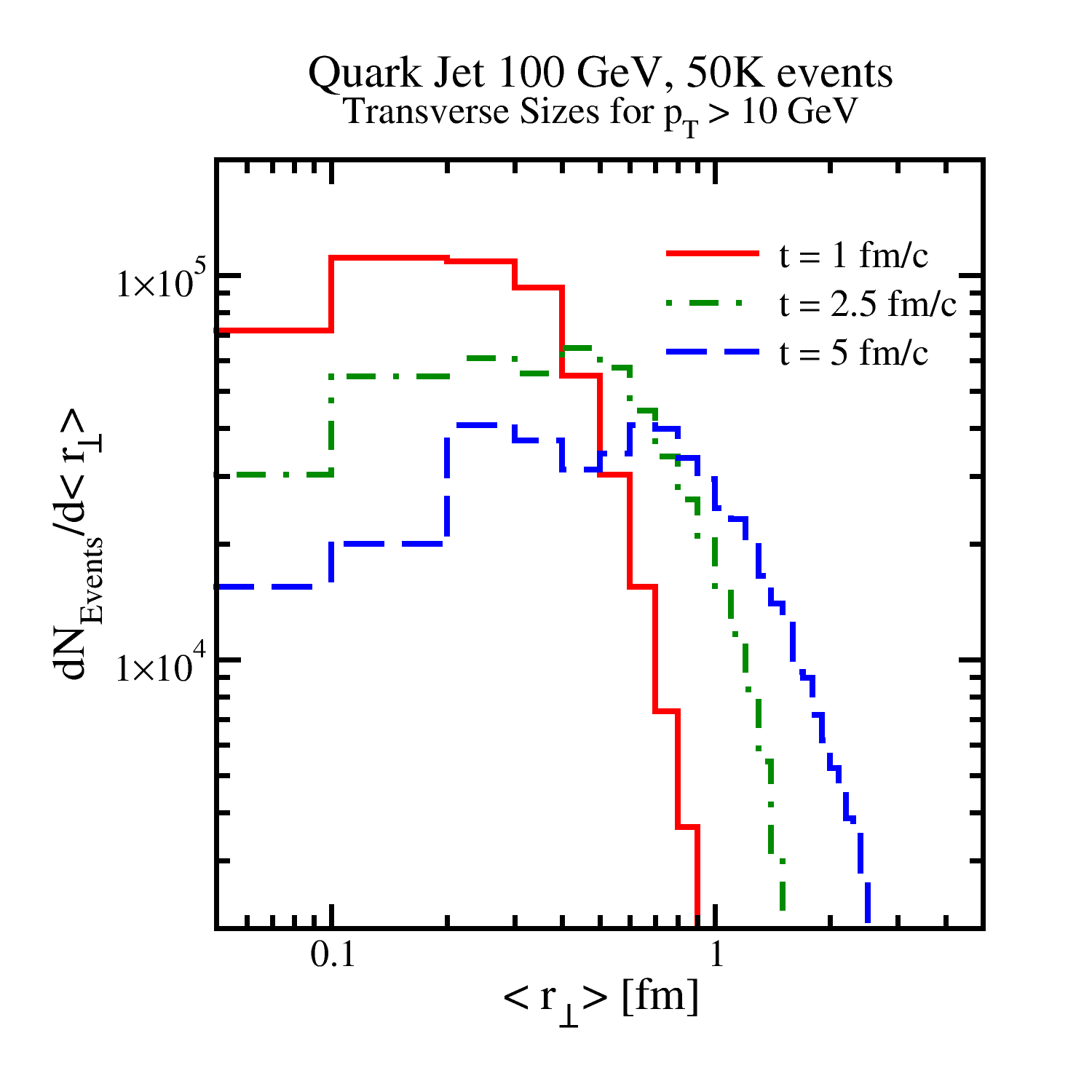}
    \caption{The dependence of the transverse size for the hard part (parton $p_T > 10$~GeV) of a jet-shower as a function of time for 3 different times: 1~fm/c (red solid), 2.5~fm/c (green dot-dashed) and 5~fm/c (blue dashed). The reader should note the log-log nature of the plot. The blue dashed 5~fm/c line is identical to the blue dashed line in Fig.~\ref{fig:size_vs_pT}.}
    \label{fig:size_vs_time_hard}
\end{figure}

One notes a gradual narrowing of the shower at $5$~fm/c, as the focus shifts to the harder core of the shower. 
However, at $5$~fm/c, the shower does not possess a \emph{perturbative} size ($r_\perp \lesssim 0.2 \,\,{\rm fm} \simeq 1$~GeV$^{-1}$). 
Its mean size is closer to $r_\perp \sim 1$~fm. 
Thus, even the hardest part of the shower is quite extended around a $t=5$~fm/c. 
The transverse size at earlier times is indeed smaller, as shown in Fig.~\ref{fig:size_vs_time_hard}. 
Here we plot the hard sector of the jet shower (with all constituent parton $p_T > 10$~GeV), at various time intervals: The red solid line is after 1~fm/c, the green dot-dashed line is after 2.5~fm/c, and the blue dashed line after 5~fm/c is the same as the blue dashed line in Fig.~\ref{fig:size_vs_pT}. 
Thus, at very early times ($1-2$ fm/c), the jet transverse size is indeed at a perturbative scale ($0.1 - 0.2$~fm). 

For the remainder of this section, we will imagine that after formation, the jet propagates through a static quark-gluon plasma (QGP) with a $T\gtrsim 0.3$~GeV, and then return to the case of jets propagating through a large nucleus (in A-DIS). The two sets of numerical values that follow, illustrate the different abilities of cold-confined and hot-deconfined, strongly interacting media to affect the hard core of jets.

Typical arguments based on \emph{coherence} or interference in jet quenching~\cite{Mehtar-Tani:2011hma,Casalderrey-Solana:2011ule, Casalderrey-Solana:2012evi} indicate that the early hard portion of the jet is too small to be resolved by typical excitations in the medium. This depends on the scale of the medium.  While this may be true at $t = 1$~fm/c, it is hard to see this by 2.5~fm/c. One should recall that in most modern simulations of the bulk evolution, the QGP is not even fully formed at 1~fm/c~\cite{JETSCAPE:2020shq,JETSCAPE:2020mzn}. A typical thermal scale in a QGP at RHIC or LHC (at $T\simeq 0.3 - 0.4$~GeV) is $\lambda_T \sim \sfrac{\hbar}{T} \sim 0.5 - 0.67$~fm. Looking at the plot in Fig.~\ref{fig:size_vs_time_hard}, it is somewhat hard to accept that there is no interaction with the medium by 2.5~fm/c. This argument for no interaction with the medium, at early times, becomes further tenuous as one uses the harder medium scale of $\mu_{\rm med}^2 = \hat{q} t$. For the case of nucleus-nucleus ($A$-$A$) collisions, we use the estimate of a static QGP, where at $t=1$~fm/c, 
\begin{align}
\lambda_\mu^{AA}(t=1{\rm fm/c})  \simeq \frac{1}{ \mu_{\rm med}} \simeq \frac{1}{ \sqrt{\hat{q} t } } \simeq 0.2\, {\rm fm}.    \label{eq:lambda-AA}
\end{align}
The last equality is obtained by setting $\hat{q} = 1$~GeV$^2$/fm (typical of fits for jet quenching in a QGP at RHIC and LHC~\cite{Majumder:2011uk}). These sizes can be easily achieved by jet showers at $t=1$~fm/c. Thus, one should assume that a QGP medium can easily resolve the hard core of jets already at 1~fm/c. 

Returning to the case of quark jets in a large nucleus (A-DIS), we can obtain the $\hat{q}$ from fits of the same formalism to the attenuation of leading hadrons as a function of their momentum fraction $z=p_h^-/q^-$~\cite{Majumder:2009zu,Majumder:2004pt}. For a typical value of $\hat{q} \approx 0.1$~GeV$^2$/fm, we obtain the resolution scale of the cold medium in a large nucleus at $t=2.5$~fm/c as, 
\begin{equation}
   \lambda_\mu^{eA} (t=2.5{\rm fm/c})  \simeq \frac{1}{ \mu_{\rm med}} \simeq \frac{1}{ \sqrt{\hat{q} t} } \simeq 0.4\, {\rm fm}. 
\end{equation}
Our simulation results in Fig.~\ref{fig:size_vs_time_hard}, show that the jet easily grows past a transverse radius of $0.4$~fm after $t=2.5$~fm/c, even in the vacuum.
The rather weak dependence on $\hat{q}$ is due to the square root taken in the equations above. 
Thus, the hard core of jets is also not un-resolved by the cold confined medium of a large nucleus after 2.5~fm/c. This does not mean that the interaction with the medium is strong, at this time. 
These observations simply argue for the need of a more developed formalism for jet medium interaction, during the early high-virtuality stage of the jet. The subsequent section will specifically address this.

In the preceding section, we derived the transverse size of a splitting parton in vacuum, from quantum uncertainty arguments. 
In this section, we provided a semi-classical or semi-quantum visualization of this uncertainty size, e.g., in Fig.~\ref{fig:uncertainty-size}, and incorporating this in Monte-Carlo simulations, generated the transverse size of developing jet showers, still entirely in the vacuum. 
The generated transverse sizes strongly question the \emph{coherence} based arguments that the early, high virtuality jet shower is too small to be resolved by the medium. 
We repeat again, this does not mean that the interaction with the medium at this stage is strong. 
In the subsequent section, we calculate the waxing of this interaction with the medium at higher parton virtuality.

\section{Medium Modified Kernel from DIS}
\label{sec:Kernel}

In the preceding sections, we first calculated the transverse uncertainty size of a splitting quark, produced in a hard interaction. Using this iteratively in a parton shower, we then demonstrated that jet showers are rather broad objects, with a non-trivial dependence of the transverse size on the initial virtuality $Q$ of the hard parton. We further demonstrated that, similar to a single split, showers get broader with time. However, even at times as early as 1~fm/c, the typical shower from a 100~GeV quark still seems to be broad enough to be resolved at the medium scale of $\mu_{\rm med}^2 \approx \sqrt{2 \hat{q} E}$ (assuming $\hat{q} \sim 1$~GeV$^2$/fm).

We reiterate: Simply because the jet shower is resolved does not mean that the interaction with the medium is strong. 
In the high virtuality stage of a jet, the effect of the medium on a split is quantified by the ratio, 
\begin{align}
 \frac{\hat{q} \tau}{\mu^2} \simeq \frac{\hat{q} 2E}{\mu^4} \ll 1,
\end{align}
where, $E$ and $\mu$ are the energy and virtuality of a given parton in the shower that is about to split, and $\hat{q}$ is the transport coefficient specific to that parton. 
Beyond this, there is still a further weakening of the effective transport coefficient at large virtualities, i.e., 
there is an energy and scale dependence in $\hat{q}$: 
\begin{align}
    \hat{q} \rightarrow \hat{q}(E, \mu^2).
\end{align}
In this section, we will evaluate this dependence. 

To evaluate the effective dependence of $\hat{q}$ on the energy $E$ and virtuality $\mu^2$ of the hard parton, we will return to the straightforward process of deep-inelastic scattering on a large nucleus, as in Sec.~\ref{sec:size_of_vacuum_emissions}.
The hadronic tensor, in the case of single gluon emission and scattering in the final state is given as, 
\begin{eqnarray}
\label{Eqn:Diff_HadTensor}
 \mathcal{W}^{\mu \nu }  \!\!& = &\!\! \sum_q\! \int \! \frac{dy d^2 \ell_\perp}{\pi} f_q^A (x) 
\frac{ d\mathcal{ K } }{d\ell_\perp^2} \left( q^-, y ,\ell_\perp^2\right) H_0^{ \mu\nu } ,
\end{eqnarray}
where $H_0^{\mu\nu}$ is the partonic hard part \cite{Guo:2000nz,Wang:2001ifa,Sirimanna:2021sqx} and $\mathcal{ K } \left( q^-, y \right)$ is known as the medium modified kernel, and $f^A_q(x)$ is the nuclear parton distribution function of a quark with momentum fraction $x$. The nucleus moves in the ($+$)-light cone direction with momentum $P_A^+ = A P^+$. The virtual photon has momentum components $q\equiv (q^+,q^-,0,0)$, with $q^+ = -x_B P^+$, where $x_B = Q^2/(2P^+ q^-)$ is the Bjorken-$x$ variable.
In the equation above, $yq^-, \ell_\perp$ are the large light-cone and transverse momentum components of the  radiated gluon.
Note that $x$ in $f_q^A(x)$ is a function of $x_B, y, \ell_\perp^2$.

The expressions above appear different from those in Sec.~\ref{sec:size_of_vacuum_emissions}. In Sec.~\ref{sec:size_of_vacuum_emissions}, we did not consider scattering in the final state. We were interested in elucidating the transverse size of the radiating hard quark. In the current section, we are interested in studying the momentum structure of the single gluon emission kernel at next-to-leading twist, that leads to medium modified emission (or stimulated emission). As a result, all of the position integrations that were analysed in Sec.~\ref{sec:size_of_vacuum_emissions}, have been integrated out.
The factor $y$ is the light-cone momentum fraction of the radiated gluon and not a position, as in Sec.~\ref{sec:size_of_vacuum_emissions}.

\begin{figure}
    \centering
    \includegraphics[width=0.4\textwidth]{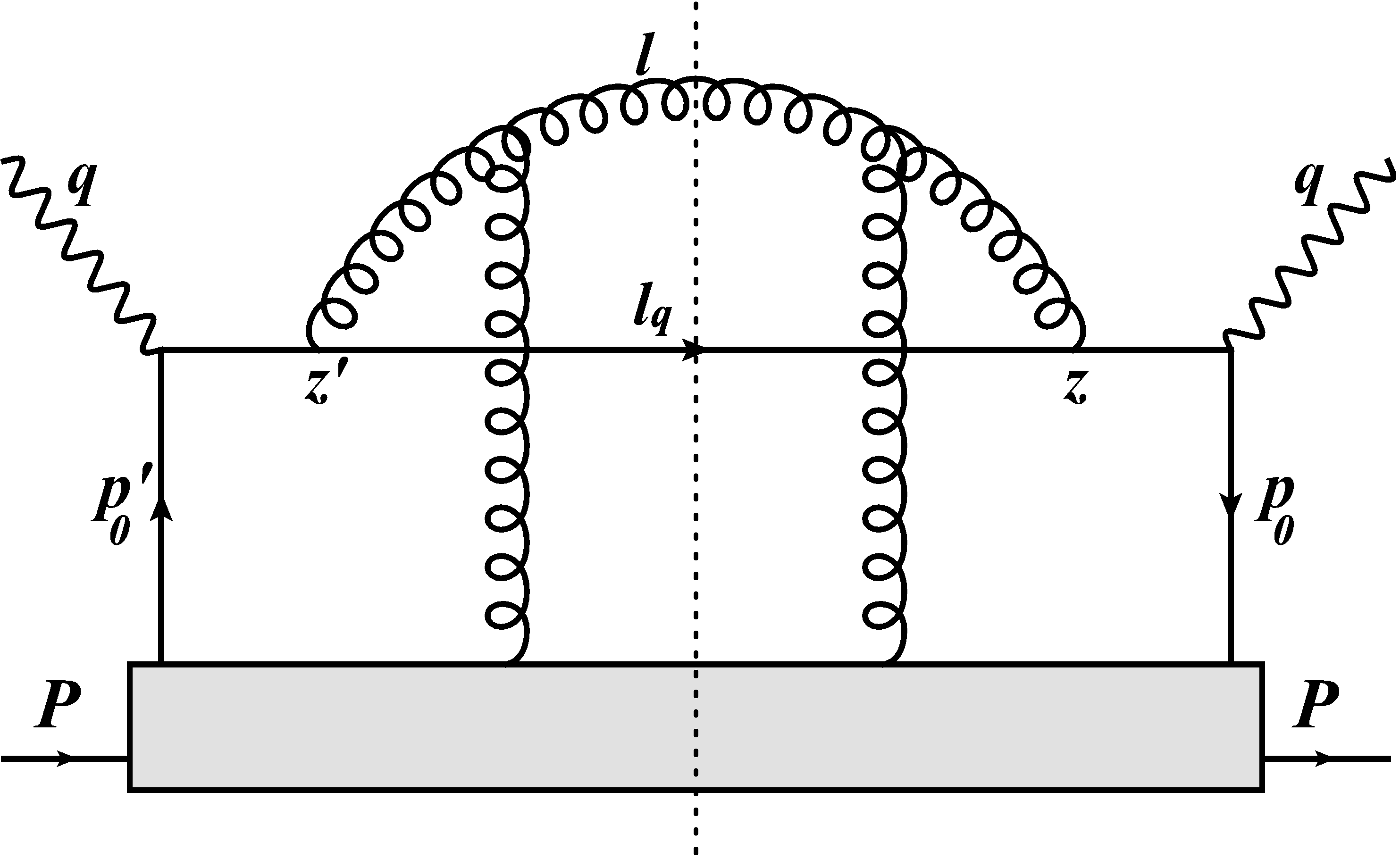}
    \caption{Single gluon emission from the outgoing struck quark produced in the deep-inelastic scattering off a large nucleus followed by a scattering on gluon. This is the only contributing Feynman diagram when $k_\perp$ dependence in the phase is ignored in the $k_\perp$ expansion. }
    \label{fig:higher-twist-diagram}
\end{figure}

\subsection{Medium Modified Kernel in the limit of soft gluon emission approximation}
\label{subsec:GWKernel}

In Eq.~\eqref{Eqn:Diff_HadTensor} the gluon emission kernel from the hard outgoing quark is medium modified. The medium modification arises from the scattering of this hard quark and radiated gluon off the glue field of the remaining nucleons, through which the quark and gluon propagate prior to hadronization. The excess radiation due to the scattering is denoted as energy loss of the fragmenting quark.

Parton energy loss in DIS is calculated within the soft and collinear gluon emission approximation using the higher-twist formalism in Refs. \cite{Guo:2000nz, Wang:2001ifa}. This involves several diagrams. However, if the factors of $k_\perp$ in the phases (where they are divided by $q^-$) are ignored in the later Taylor expansion in $k_\perp$, then only one diagram contributes: The double gluon scattering diagram in Fig.~\ref{fig:higher-twist-diagram}.  Incorporating this, the medium-modified kernel before collinear expansion is given by
%
%
\begin{align}
  \mbox{} &  \mathcal{K}_\mathrm{GW}  \left( q^-, y \right) = \frac{\alpha_s}{ 2 \pi } P(y)\int d^2 \ell_\perp 
    \int \frac{ d \xi^- d \delta \xi^- d^2 k_\perp d^2 \xi_\perp}{ ( 2 \pi )^2 }  \nonumber\\
    \times &  \frac{ \left\{ 2 - 2 \cos \left( \frac{ \left( \vec{\ell}_\perp - \vec{k}_\perp \right)^2 \xi^- }{ 2 q^- y \left( 1 - y \right) } \right) \right\} }{ \left( \vec{ \ell }_\perp - \vec{ k }_\perp \right)^2 } e^{-i \frac{\ell_\perp^2}{2 q^- y (1 - y)} \delta \xi^-}  \label{Eqn:GW_kernel} \\
    \times & e^{-i \frac{ k_\perp^2 - 2 \vec{k}_\perp \cdot \vec{\ell}_\perp }{ 2 y q^- } \delta \xi^- + i \vec{k}_\perp \cdot \vec{\xi}_\perp} \nonumber \\
   \times &  \bra{A} A^{+} \left(\xi^- + \delta \xi^-, \vec{\xi}_\perp \right) A^+ \left( \xi^-, 0 \right) \ket{A}. \nonumber
\end{align}
In the equation above, $\xi^-, \delta \xi^-$ represent the mean and offset in the light-cone locations of the scattering in the amplitude and complex conjugate, as depicted in Fig.~\ref{fig:higher-twist-diagram}, and 
$\xi_\perp$ is the offset in the perpendicular locations. 
The above expression is derived in the light-cone gauge $A^- = 0$. In this gauge, $A^+ \gg A_\perp$~\cite{Majumder:2009ge}.

The differential yield of induced gluons, when the medium interaction is within the single gluon scattering per emission limit can be given by,
\begin{align}
   \mbox{} & \frac{dN_g}{dy d\ell_\perp^2} =  \frac{d \mathcal{K}_\mathrm{GW}}{d\ell_\perp^2} = \frac{\alpha_s}{2\pi^2} P(y) \int \frac{ d \xi^- d \delta \xi^- d^2 k_\perp d^2 \xi_\perp}{ ( 2 \pi )^2 } \nonumber\\
    \times & \frac{ \left[ 2 - 2 \cos \left\{ \frac{ \left( \vec{\ell}_\perp - \vec{k}_\perp \right)^2 \xi^- }{ 2 q^- y \left( 1 - y \right) } \right\} \right] }{ \left( \vec{ \ell }_\perp - \vec{ k }_\perp \right)^2 } 
    e^{-i \frac{ (\vec{\ell}_\perp - \vec{k}_\perp)^2 }{ 2 y q^- } \delta \xi^-}   \label{Eqn:GluonSpectrum_GW} \\
    \times & e^{-i \frac{\ell_\perp^2}{2 q^- (1 - y)} \delta \xi^-} e^{i \vec{k}_\perp \cdot \vec{\xi}_\perp} \bra{A}A^{+} \left(\delta \xi^-, \vec{\xi}_\perp \right) A^+ \left( 0 \right) \ket{A} . \nonumber
 \end{align}
One immediately notes that the $\xi^-$ and $\delta \xi^-$ integrations are factorized in the expression above. For the specific case of a system with translation invariance, i.e., the $\langle A^+ A^+ \rangle$ 
correlator does not depend on the mean location $\xi^-$ (only on the relative separation), we can drop the $\xi^-$ dependence in the correlator. The $\xi^-$ integration can now be completed, integrating from $\xi^- = 0$ to $\xi^- = 2\tau_f^- = \frac{4q^-y(1-y)}{l_\perp^2}$, 
\begin{align}
    \mbox{} & \frac{dN_g}{dy d\ell_\perp^2} = \frac{\alpha_s}{2\pi^2} P(y) \int \frac{d^2 k_\perp }{(2\pi)^2} \label{eqn:time-integrated-gluon-spectrum} \\
    & \times \frac{ \frac{8q^-y(1-y)}{\ell_\perp^2}  
    - \frac{4 q^- y(1-y)}{( \vec{\ell}_\perp - \vec{k}_\perp )^2 } 
    \sin\left\{  \frac{2 \left( \vec{ \ell }_\perp - \vec{ k }_\perp \right)^2 }{\ell_\perp^2 }  \right\}    }
    { \left( \vec{ \ell }_\perp - \vec{ k }_\perp \right)^2 } \nonumber \\
    &\times \int  d \delta \xi^-  d^2 \xi_\perp 
     e^{-i \frac{ (\vec{\ell}_\perp - \vec{k}_\perp)^2 }{ 2 y q^- } \delta \xi^-}  
     e^{-i \frac{\ell_\perp^2}{2 q^- (1 - y)} \delta \xi^-} e^{i \vec{k}_\perp \cdot \vec{\xi}_\perp} \nonumber \\
     & \times \bra{A}A^{+} \left(\delta \xi^-, \vec{\xi}_\perp \right) A^+ \left( 0 \right) \ket{A} .
     \nonumber
\end{align}

\begin{figure}
    \centering
    \includegraphics[width=0.45\textwidth]{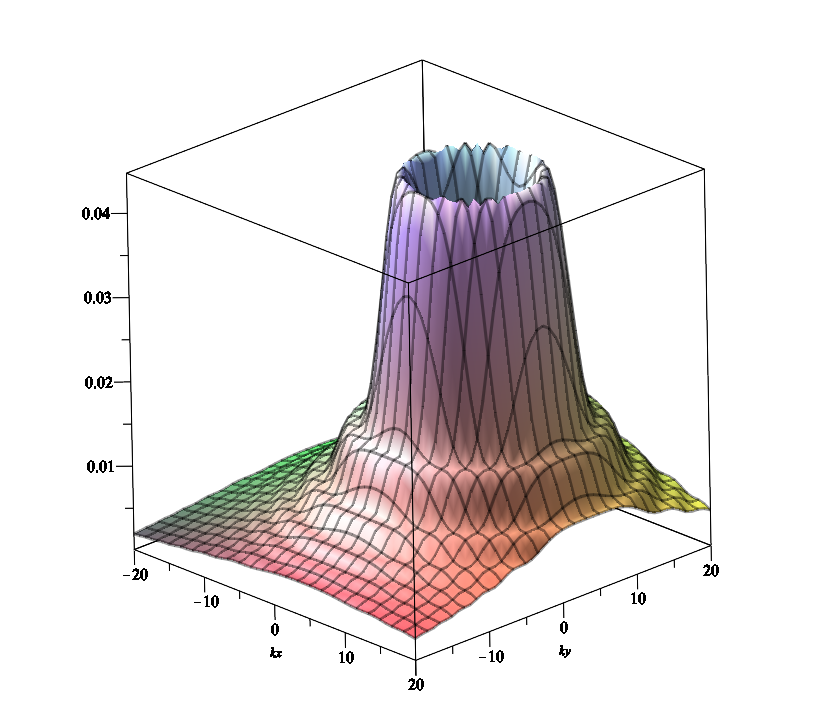}
    \caption{(Color Online) A plot of the hard portion (2nd line) of the differential gluon spectrum $\sfrac{dN_g}{dy d \ell_\perp^2}$ in Eq.~\eqref{eqn:time-integrated-gluon-spectrum} as a function of $\vec{k}_\perp = k_x \hat{x} + k_y \hat{y}$. For the plot, $q^- \!\!= 100$~GeV and $y=0.1$. Also we have set $\ell_x \!=\! \ell_y\! =\! 4$~GeV ($\vec{\ell}_\perp = \ell_x \hat{x} + \ell_y \hat{y}$). The function yields a steep cone like structure centered at $\vec{k}_\perp = \vec{\ell}_\perp$, with a rim radius approximately equal to $| \vec{\ell}_\perp |$.}
    \label{fig:h_time_integrated}
\end{figure}

The general form of the equation above may be expressed as,
\begin{align}
    \frac{dN}{dy d\ell_\bot^2 } &= \frac{\alpha_S P(y)}{\pi} \int d^2 k_\perp \nonumber \\
    &\times h (\vec{k}_\perp,q^-,y,\vec{\ell}_\perp) \times g (\vec{k}_\perp,q^-,y,\vec{\ell}_\perp),
\end{align}
where, $h$ contains the hard dynamics involving $\ell_\perp$, $k_\perp$ and $q^-$, given by the second line of Eq.~\eqref{eqn:time-integrated-gluon-spectrum}, and $g$
represents the soft matrix element (third and fourth line of the equation).
A plot of $h$ for $q^-=100$~GeV, $y=0.1$, and $\ell_x = \ell_y = 4$~GeV is shown in Fig.~\ref{fig:h_time_integrated}. The function peaks in a conical structure around the point $k_\perp = \ell_\perp$ (in fact, $h=0$ at $k_\perp = \ell_\perp$). The rim-like structure has a radius of $\delta k_\perp \simeq |\vec{\ell}_\perp|$. Thus, the preponderance of values of $k_\perp$ that enhance $h$ is large and of the order of $|\vec{\ell}_\perp|$. This is in contrast with the soft matrix element $g$, which is enhanced by small values of $k_\perp$, typically in the range $\sim 1$~GeV. In this paper, we are interested solely in the large virtuality limit, where $\ell_\perp \gg 1$~GeV. 

\subsection{The Soft Matrix Element}
\label{subsec:matrix-element}

Taking the limit of a large $\ell_\perp$, which is much larger than the eventual $k_\perp$, we may simplify the equation for the soft part $g(\vec{k}_\perp,q^-,\vec{\ell}_\perp)$ as, 
\begin{align}
    g(k_\perp,q^-,\ell_\perp) &\simeq \int \frac{d\delta \xi^- d^2 \xi_\perp }{ (2\pi)^3  }
    e^{-i \frac{l_\perp^2}{2q^-y(1-y)} \delta \xi^- } e^{i \vec{k}_\perp \cdot \vec{\xi}_\perp}
    \nonumber \\
    &\times \bra{A} A^+ (\delta\xi^-, \vec{\xi}_\perp ) A^+ (0,0) \ket{A}.
\end{align}

Dropping the $\vec{k}_\perp$ in the exponential for $\delta \xi^-$, along with the assumption that the two-point function is symmetric in $\phi_{\xi_\perp}$, leads to $g(k_\perp, q^-,y, \ell_\perp)$ becoming independent of the azimuthal angle for $k_\perp$. The factor $g(k_\perp, q^-,y, \ell_\perp)$, can then be expressed as, 
\begin{align}
    g(k_\perp, &q^-,y,\ell_\perp) = \frac{ x_L P^+ }{4 k_\perp^2} 
    \int \frac{d\delta \xi^-  d^2 \xi_\perp }{(2\pi)^3} 
    \frac{e^{-ix_L P^+ \delta \xi^-} e^{i \vec{k}_\perp \cdot \vec{\xi}_\bot} }{x_L P^+} \nonumber \\
    &\times \bra{A} F^{+ \bot} \left(\frac{\delta\xi^-}{2}, \frac{\vec{\xi}_\perp}{2} \right) 
    F_{\bot , }^{+} \left(\frac{-\delta\xi^-}{2}, -\frac{\vec{\xi}_\perp}{2} \right)  \ket{A} \nonumber \\
    &= \frac{ \rho x_L  }{8 k_\perp^2} \int \frac{d\delta \xi^-  d^2 \xi_\perp }{(2\pi)^3} 
    \frac{e^{-ix_L P^+ \delta \xi^-} e^{i \vec{k}_\perp \cdot \vec{\xi}_\bot} }{x_L P^+} \nonumber \\
    &\times \bra{P} F^{+ \bot} \left(\frac{\delta\xi^-}{2}, \frac{\vec{\xi}_\perp}{2} \right) 
    F_{\bot , }^{+} \left(\frac{-\delta\xi^-}{2}, -\frac{\vec{\xi}_\perp}{2} \right)  \ket{P} \nonumber \\
    &= \frac{\rho x_L  }{8 k_\perp^2}  G(x_L, k_\perp). 
\end{align}
In the equation above, $x_L = \sfrac{\ell_\perp^2}{2q^-P^+ y(1-y)}$, where $P^+ = P_A^+/A$ is the average lightcone momentum of one nucleon in the nucleus. We have further decomposed the nuclear state ($\ket{A}$) into nucleon states ($\ket{P}$) as, 
\begin{align}
    \bra{A} \ldots \ket{A} &\simeq \frac{\rho}{2P^+} \bra{P} \ldots \ket{P},
\end{align}
where $\rho$ is the density of nucleons within the nucleus.
The factor $G(x_L, k_\perp)$ represents the gluon TMDPDF, and we have assumed that the operator product is translation invariant within the nuclear state $\ket{A}$. TMDPDFs even at leading order have a scale ($\bar{\mu}$) dependence~\cite{Aybat:2011zv}. In this case, the scale of the TMDPDF is obviously $\bar{\mu}^2 \approx \ell_\perp^2$. Thus, we will evaluate the quantity $G(x_L,k_\perp,\ell_\perp^2)$.  One notes that $\bar{\mu}^2 \simeq \mu^2$, the virtuality of the hard parton. Thus, in the remainder of this effort, we will simply use $\mu^2$ as the scale of the TMDPDF, and hence, the eventual scale of the transport coefficient $\hat{q}$.

\subsection{Relating $\hat{q}$ to the TMDPDF}
\label{subsec:q-hat-from-TMDPDF}

The entire expression to be evaluated is, 
\begin{align}
     \frac{dN}{dy d\ell_\bot^2 } &= \frac{\alpha_S P(y)}{\pi}\int d^2 k_\perp 
     h(\vec{k}_\perp, q^-,y, \vec{\ell}_\perp ) \nonumber \\
     &\times \frac{  \rho x_L }{8 k_\perp^2}   G (x_L, \vec{k}_\perp,\ell_\perp^2),
     \label{eq:gluon_yield_depending_on_TMDPDF}
\end{align}
The standard method for calculating the above expression is to Taylor expand $h(\vec{k_\perp}, q^-, \vec{\ell}_\perp )$ in $\vec{k}_\perp$, and take the second derivative, reducing the gluon yield to, 
\begin{align}
    \frac{dN}{dy d\ell_\bot^2 } &= \frac{\alpha_S P(y)}{\pi}\int d^2 k_\perp 
     \left. \frac{\vec{\nabla}^2_{k_\perp} h(\vec{k}_\perp, q^-, y, \vec{\ell}_\perp ) }{2!} \right|_{k_\perp =0} \nonumber \\
     &\times \frac{ \rho x_L  }{8 }   G (x_L, \vec{k}_\perp,\ell^2_\perp).
\end{align}
The above expression is often expressed in terms of the jet transport coefficient $\hat{q}$ as, 
\begin{equation}
   \mbox{}\!\!\! \frac{dN}{dy d\ell_\bot^2 } \!=\! 
   \frac{\alpha_S P(y)}{2 \pi}\!\! 
   \left.\vec{\nabla}^2_{k_\perp} \! 
   h(\vec{k}_\perp, q^-\!, y, \vec{\ell}_\perp ) \right|_{k_{\perp} = 0} \!\!
   \hat{q} (x_L,\ell^2_\perp). \label{eq:gluon_yield_depending_on_qhat}
\end{equation}
In the above equation, $\hat{q}$, given as $\int d^2 k_\perp \rho x_L G(x_L, \vec{k}_\perp, \ell_\perp^2 )/8$, has an obvious dependence on scale $(\ell_\perp^2)$ and the energy of the hard parton ($q^-$), the momentum fraction of the radiated gluon ($y$), via the momentum fraction $x_L = \ell_\perp^2/[ 2 q^- y (1-y)]$. It also may depend on location; we are neglecting this dependence by assuming $\rho$ to be a constant. Using Eq.~\eqref{eq:z-with-l_perp}, the dependence on scale can be thought of as a dependence on the transverse size ($z_\perp$) of the radiating antenna. We continue to use terminology such as ``antenna", for tha lack of a better term, even though our calculations refer to configurations such as those in Fig.~\ref{fig:uncertainty-size}.

The above discussion implies that the interaction of a radiating parton with a medium (qualified by $\hat{q}$) depends on, among other things, the size of the radiating antenna (Fig.~\ref{fig:uncertainty-size}). The above discussion is somewhat different than the usual arguments of jet coherence, as it considers the transverse size at the point of formation and not after formation. Although, its overall phenomenological effect~\cite{JETSCAPE:2022jer,JETSCAPE:2024cqe} may indeed be indistinguishable from phenomenology based on typical coherence arguments. 

Thus, to obtain the full scale (size) dependence of $\hat{q}$, we simply compare Eq.~\eqref{eq:gluon_yield_depending_on_TMDPDF} with Eq.~\eqref{eq:gluon_yield_depending_on_qhat}, to obtain,
\begin{align}
    \hat{q}(x_L,\ell_\perp^2 ) \!=\! \frac{ \int\! \frac{d^2 k_\perp \rho x_L}{4 k_\perp^2} h(\vec{k}_\perp, q^-\!,y, \vec{\ell}_\perp \!) 
     G (x_L, \vec{k}_\perp,\ell_\perp^2\!)}
    { \left.\vec{\nabla}^2_{k_\perp} \! 
   h(\vec{k}_\perp, q^-\!, y, \vec{\ell}_\perp ) \right|_{k_{\perp} = 0}  }. \label{eq:qhat_TMDPDF}
\end{align}
The equation above, as written, is more general than the higher-twist approach used to derive it. It directly relates the TMDPDF of a given medium with the transverse momentum coefficient $\hat{q}$, and is true for any choice of $h$. Similar relations have been suggested before~\cite{Wang:2001ifa,Kumar:2019uvu}.
However, in this effort, we provide a straightforward derivation of this relation. It demonstrates with some clarity that the scale dependence of $\hat{q}$ is directly related to the transverse momentum $k_\perp$, and the scale and $x$ dependence of the TMDPDF within the medium.

\section{Results on Scale dependence of $\hat{q}$}
\label{sec:Results}

In the preceeding section, we derived a relation between the gluon TMDPDF within the nucleons of a large nucleus and the jet transport coefficient $\hat{q}$.
Eq.~\eqref{eq:qhat_TMDPDF} can be solved at varying levels of complexity. The $h$-function depends on approximations used in the derivation of the high-virtuality energy-loss kernel. The TMDPDF can be evaluated at varying levels of sophistication. In this paper, we will present the simplest possible version of this equation. The goal is for these results to be easily reproduced. More realistic calculations, including the attempt to extend these results to jets in a QGP will be left for a future effort. Our goal here is to simply elucidate the mechanics of the effect, making a clear connection between the TMDPDF and the scale dependence of $\hat{q}$.

As such, we will use the method of Ref.~\cite{Wang:2001ifa}, to obtain $\left.\vec{\nabla}^2_{k_\perp} h(\vec{k}_\perp, q^-\!, y, \vec{\ell}_\perp ) \right|_{k_{\perp} = 0}$, where one ignores derivatives on the phase. Although this was criticized in Refs.~\cite{Aurenche:2008hm,Aurenche:2008mq} for being incomplete, a more general analysis including the full set of terms at next-to-leading twist~\cite{Sirimanna:2021sqx}, showed that the method of Ref.~\cite{Wang:2001ifa} is actually close to the complete result at $\tau = 2\tau_f$. Within this approximation, we obtain,
\begin{align}
   &\mbox{}\!\!\!\! \left.\vec{\nabla}^2_{k_\perp}\!\! h(\vec{k}_\perp, q^-\!, y, \vec{\ell}_\perp ) \right|_{k_{\perp} \!\!= 0} \!\!\!= \!\frac { 16 q^-\!y\!\left( 1\!-\!y \right) \pi\! \left[ 2\!-\!\sin \left( 2 \right)
 \right] }{\ell_\perp^{6}}.
\end{align}
Note that the corresponding $h$ factor is given by the second line of Eq.~\eqref{eqn:time-integrated-gluon-spectrum}. 

We follow Aybat and Rogers~\cite{Aybat:2011zv}, and express the gluon TMDPDF, at lowest order as, 
\begin{align}
    G(x,\vec{k}_\perp , {\mu}^2) &= F(x) \frac{{\rm e}^{-\frac{k_\perp^2}{4w(x,{\mu}^2)}} }{ 4 \pi w (x,{\mu}^2)}, \label{eq:gluon_TMDPDF}
\end{align}
where, $F(x)$ is the regular collinear parton distribution function (PDF) at lowest order, and the transverse width $w$ is given in the Brock-Landry-Nadolsky-Yuan form~\cite{Landry:2002ix}: 
\begin{equation}
   w(x,{\mu}^2) =  \frac{ g_2 \log\left( \frac{{\mu}}{2Q_0} \right)}{2} + g_1 \left[\frac{1}{2} + g_3 \log(10x) \right].
\end{equation}
The constants in the above equation are $Q_0 = 1.6$~GeV, $g_1 = 0.21$~GeV$^2$, $g_2 = 0.68$~GeV$^2$, $g_3 = 1.6$. Following Ref.~\cite{Aybat:2011zv}, we take $F(x)$ from the Martin-Stirling-Thorne-Watt (MSTW) leading order parametrization, given in Ref.~\cite{Martin:2009iq} as,
\begin{equation}
    F(x) = A x^\delta (1-x)^\eta (1 + \epsilon\sqrt(x) + gx),
\end{equation}
with $A=0.0012216$, $\delta = -1.83657 $, $\eta = 2.3882$, $\epsilon = -38.997$ and $g=1445.5$. Including higher orders will require evolving the above TMDPDF using the Collins-Soper equation~\cite{Collins:1981uw}.
In the interest of simplicity and ease of reproducibility, we will refrain from this, and use the TMDPDF as written in Eq.~\eqref{eq:gluon_TMDPDF}.

In a realistic calculation, Eq.~\eqref{eq:gluon_TMDPDF} provides the initial condition at a low value of ${\mu}^2$, and has to be evolved using the Collins-Soper equation to a ${\mu}^2$ that is relevant to the physical problem. Carrying out this evolution will involve a certain amount of numerical computation. It will lead to a quantitative, but \emph{not a qualitative}, shift in our results. In a future effort, we plan to carry out this calculation using the full form of the hard sector function $h(\vec{k}_\perp,q^-,y,\vec{\ell}_\perp)$, and an evolved form of the TMDPDF. The results presented in this paper can be reproduced by anyone, either by hand, or using analytic manipulation software such as \emph{Mathematica} (without the explicit use of any advanced numerical techniques). 

To carry out the $d^2 k_\perp$ integration in the numerator of Eq.~\eqref{eq:qhat_TMDPDF}, requires an infrared regulator; we have set the minimum value of $k_\perp$ to be $\Lambda_{QCD} = 0.2$~GeV. Incorporating all these terms, we can obtain the effective $\hat{q}$ as a function of the inverse antenna size, given by $Q^2 \simeq \ell_\perp^2$ at $x_L = \sfrac{\ell_\perp^2}{[2 q^- P^+ y (1-y)]}$. For this estimate, since all our expressions are boost invariant for boosts in the $z$-direction, we boost to the target frame and pick $q^- = 200\sqrt{2}$~GeV, with $P^+ = \sfrac{1}{\sqrt{2}}$~GeV. We pick the radiated gluon's momentum fraction as $y=\sfrac{1}{2}$. In order to highlight the effect of evolution, we follow Ref.~\cite{JETSCAPE:2022jer}, and write 
\begin{equation}
    \hat{q}(\mu^2) = \hat{q} \cdot f(\mu^2), \label{eq:qhat-f}
\end{equation}
i.e., we place the entire scale dependence in $f(\mu^2)$.

Given the form of the TMDPDF, values of $\mu= \ell_\perp \equiv  \sqrt{ ( \vec{\ell}_\perp)^{\,2} } \lessapprox 3$~GeV correspond to a significant narrowing of the width $w(x_L,\mu^2)$, coupled with a sharp rise in $F(x_L)$  [note $x_L = \sfrac{\ell_\perp^2}{2q^-P^+ y(1-y)}$]. As a result, given the choice of parameters, values of $\mu\lessapprox 3$~GeV require a more careful numerical treatment. On the opposite end, for a $q^- = 200\sqrt{2}$ and $y=\sfrac{1}{2}$, the maximum allowed value of $\ell_\perp = 10$~GeV. Beyond this, the collinear PDF vanishies, i.e. $F(x_L) \rightarrow 0$. As a result, we present the evolution of $\hat{q}$ with scale $\mu(=\ell_\perp)$ from 3.1~GeV to 9~GeV.

\begin{figure}
    \centering
    \includegraphics[width=0.9\linewidth]{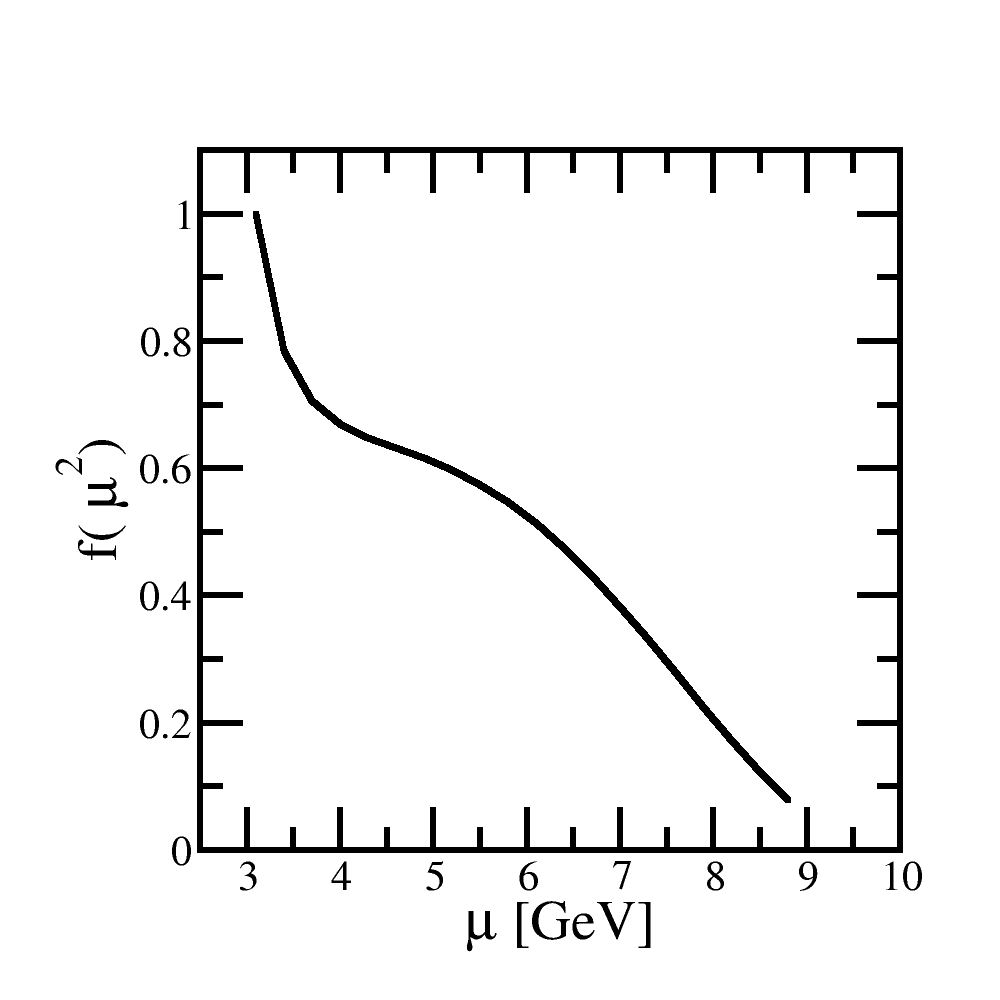}
    \caption{A plot of the scaling function for $\hat{q}$ as described in Eq.~\eqref{eq:qhat-f}. $f(\mu^2)$ is determined by solving Eq.~\eqref{eq:qhat_TMDPDF} and shows a rapid drop as a function of $\mu$. }
    \label{fig:fQ}
\end{figure}

In Fig.~\ref{fig:fQ}, we present the behavior of the scaling function $f(\mu^2)$ from $\mu=3.1$~GeV to 9~GeV. 
The function $f$ is defined such that at $\mu= 3.1$~GeV, $f=1$ and $\hat{q}$ is the normalization, i.e., the value of $\hat{q}(\mu^2) $ at $\mu=3.1$~GeV. The scaling function shows a strong dependence on the scale $\mu^2$, almost vanishing by the point $\mu \rightarrow 9$~GeV. This is entirely consistent with the experimentally extracted forms of $f(\mu^2)$ in Refs.~\cite{JETSCAPE:2022jer,JETSCAPE:2024cqe}, even though, those extractions were conducted for jets in a quark-gluon plasma. This is also consistent, though not obviously so, with the scale dependence of $\hat{q}$ extracted in comparison with experimental data at RHIC and LHC energies in Ref.~\cite{Kumar:2019uvu}. 

In prior efforts dealing with jet coherence and the non-interaction of jets of small transverse size with the medium, the entire effect is ascribed to the small size of the radiating antenna~\cite{Casalderrey-Solana:2012evi,Mehtar-Tani:2011hma,Armesto:2011ir, Casalderrey-Solana:2011ule}. However, in our calculation, which considers the effect of the medium within the formation time of a radiation, the weakening of the effective transport coefficient $\hat{q}$ with the scale of the splitting antenna arises from three separate effects. The first is the shrinking size of the radiating antenna, given by $\delta z_\perp \sim \sfrac{1}{\ell_\perp}$, which leads to the antenna being affected only by transverse momenta, from the medium, of the same size as those from the antenna, i.e. $h$ is large when $k_\perp \sim \ell_\perp$. This is illustrated in Fig.~\ref{fig:h_time_integrated}. As one increases $\ell_\perp$, it becomes less likely for the soft factor $g$ to produce a gluon with a $k_\perp \sim \ell_\perp$. This leads to the drop in $f(\mu^2)$ as $\mu$ is increased from 3~GeV. As the scale $\mu (= \ell_\perp)$ continues to increase, the width $w$ of the TMDPDF increases, and this increases the probability for larger $k_\perp$. This leads to the mild bump like structure between 4 and 6 GeV. Another reason for this structure is that it corresponds to an $x_L$ in the range from 0.16 to 0.36 which is the least steep portion of $F(x_L)$. Larger values of $l_\perp$ require comparable values of $k_\perp$, for $h\cdot g $ to stop depreciating. However, there is almost no support from the TMDPDF for values of $k_\perp$ larger that 6~GeV. This combined with the sharper drop in $F(x_L)$ for $x_L \rightarrow 1$, leads to the precipitous drop of $f(\mu^2)$ with increasing $\mu$.

A qualitative summary of the above analysis is that increasing the transverse momentum $\ell_\perp$ of a split, leading to smaller sizes of the radiating system, requires harder emissions from the medium to affect the splitting process. This becomes less probable with increasing $l_\perp$, and as a result, a radiating dipole has a progressively weaker interaction with the medium. 
This leads to a continuously decreasing effective $\hat{q}$ with increasing $l_\perp$. Unlike the coherence based arguments~\cite{Casalderrey-Solana:2012evi,Mehtar-Tani:2011hma,Casalderrey-Solana:2011ule}, there is no sudden lack of interaction with the medium as the scale of the antenna drop below the medium scale $\lambda^{AA}_\mu$ [Eq.~\eqref{eq:lambda-AA}]. Instead there is a steady decrease in the effective $\hat{q}$ due to the decreasing size of the radiating antenna, coupled with the $x$ and $k_\perp$ dependence of the gluon-TMDPDF. Hence, we refer to the gradual weakening of the interaction, as a function of the scale of the radiating antenna (and by extension the scale of the gluon TMDPDF), described in this paper, as ``Modified Coherence".

\section{Summary and outlook}
\label{sec:Summary}

Partons within jets are quantum mechanical objects. As such, their position and momentum uncertainties are related by Heisenberg's principle. For all jet observables in vacuum, this uncertainty \emph{size} is almost irrelevant. However, for jets propagating through and interacting with a dense medium, an estimate of the mean and uncertainty of the location of a parton are essential. 

In earlier efforts, the transverse size of partonic showers was considered to be small compared to the size of the medium. Jet energy-loss event generators focused only on the longitudinal location of the developing shower~\cite{Majumder:2013re,Cao:2017qpx,Schenke:2009gb,He:2015pra}. The transverse extent of the shower was either ignored or calculated using the classical antenna picture, where each parton traverses straight line trajectories.

There is now a growing consensus that the hard virtual core of jets has a diminished interaction with the medium. Phenomenological fits based on multi-stage simulations~\cite{JETSCAPE:2022jer,JETSCAPE:2023hqn,JETSCAPE:2024nkj,JETSCAPE:2022hcb,JETSCAPE:2024cqe}, which are compared with a wide swath of data, clearly indicate the presence of this diminishing interaction with increasing virtuality of the jet parton. 
Partons with a large virtuality, tend to produce spilts with a large momentum $\ell_\perp$ transverse to the direction of the parent parton. By uncertainty arguments, these splits should have a small size. This leads to the coherence based arguments that smaller antenna are not resolved by the medium and thus, these portions of hard jets evolve in the medium as a vacuum like shower~\cite{Casalderrey-Solana:2011ule,Mehtar-Tani:2011hma,Casalderrey-Solana:2012evi}. 

The above arguments have prompted the current study, which was composed of three parts. In Sec.~\ref{sec:size_of_vacuum_emissions}, we demonstrated that the transverse uncertainty size of splitting partons was indeed given by the na\"{i}ve relation $\delta z_\perp \sim 1/\ell_\perp$. In Sec.~\ref{sec:MC}, we incorporated this relation within a Monte-Carlo parton shower and demonstrated the lack of a clear relation between the virtuality of the originating parton and the transverse size of the entire shower, as well as its hard core. In spite of this, in the next section, we found that each splitting parton has a diminishing interaction with the medium as its virtuality is increased, and this is caused by 3 separate effects: the shrinking size of the radiating parton requiring a larger $k_\perp$ from the medium to affect the split, the scale dependent  transverse momentum width and the $x$ dependence of the gluon TMDPDF, which sources the medium gluon, that interacts with the hard splitting parton. 

In Sec.~\ref{sec:Kernel}, we derived a new equation [Eq.~\eqref{eq:qhat_TMDPDF}], that can be used in any jet quenching formalism to deduce the relation between a gluon TMDPDF and the jet transport coefficient $\hat{q}$.
At a phenomenological level, our results are similar to those derived using the ``coherence" approach~\cite{Mehtar-Tani:2011hma,Armesto:2011ir,Casalderrey-Solana:2011ule,Casalderrey-Solana:2012evi}. However, the exact reason for the diminishing interaction with the medium with increasing virtuality of the radiating parton is somewhat different. As a result, we refer to this effect as ``Modified Coherence". Basic and extensive phenomenology based on modified coherence, has already been carried out in Refs.~\cite{JETSCAPE:2022jer,JETSCAPE:2022hcb,JETSCAPE:2023hqn,JETSCAPE:2024cqe,JETSCAPE:2024nkj}, and show considerable agreement with a broad range of experimental data. 

Our final numerical results in Sec.~\ref{sec:Results}, were calculated using a simple model of the gluon TMDPDF, with parameters that were dialed by fits to data. We did not include a Collins-Soper evolution of the TMDPDF and explored a rather limited range in scale. Thus, our results will have to be improved upon. Our goal in this paper was to present a calculation that could be repeated by anyone without the need for any advanced numerical tools. In a future more numerical effort, one will be able to include lower scales and a full Collins-Soper evolution. Such an effort will also allow the use of the full kernel at next-to-leading twist~\cite{Sirimanna:2021sqx}.

Beyond all the enhancements mentioned above will be the effort to obtain the scaling function $f(\mu^2)$ [Eq.~\eqref{eq:qhat-f}] for a jet traversing a quark-gluon plasma (QGP). One cannot directly use Eq.~\eqref{eq:qhat_TMDPDF}, as no TMDPDF has even been defined, let alone measured, in a QGP. However, one could use Eq.~\eqref{eq:qhat_TMDPDF} in reverse: Having measured the $\hat{q}$ and its scale dependence, one could surmise the form of a TMDPDF in the QGP.

A surprising and side consequence of this effort, particularly that in Sec.~\ref{sec:MC}, is the realization that jets are rather wide objects. 
Looking at Fig.~\ref{fig:size_vs_pT}, specifically at the case with $p_T > 0$, one notes that at $t=5$~fm/c, jets have radii of several fm. 
For hard jets produced in relativistic heavy-ion collisions, this means that at least the softer portions of jets effect a much larger part of the QGP medium than previously imagined. We remind the reader that the simulations in Sec.~\ref{sec:MC} are vacuum simulations. Once repeated in the medium, these jets will only become wider. The softer part of the jets tend to dissipate in the medium leading to a wake of the jet~\cite{Tachibana:2020mtb}. Repeating these simulations, including the quantum uncertainty sizes will lead to even wider wakes of the jets.

\begin{acknowledgements}
 This work was supported in part by the U.S. Department of Energy (DOE) under grant number DE-SC0013460. and in part by the National Science Foundation (NSF) under grant number ACI-1550300 within the framework of the JETSCAPE collaboration. C.S. is supported in part by the grant DE-FG02-05ER41367 from the U.S. Department of Energy, Office of Science, Nuclear Physics.
\end{acknowledgements}

\bibliography{HigherTwistCoherence,refs}

\end{document}